\documentclass[notitlepage,a4paper,aps,prd,onecolumn,superscriptaddress,nofootinbib,groupedaddress,longbibliography]{revtex4-1}
\usepackage{amsmath}
\usepackage{amsfonts}
\usepackage{amssymb}
\usepackage{accents}
\usepackage[T1]{fontenc}
\usepackage[utf8]{inputenc}
\usepackage[colorlinks=true]{hyperref}
\usepackage[svgnames]{xcolor}
\usepackage{listings}
\usepackage[euler]{textgreek}

\newcommand{\order}[2]{\accentset{#2}{#1}}

\newcommand{\dd}{\mathrm{d}}
\newcommand{\lc}[1]{\accentset{\circ}{#1}}
\newcommand{\tp}[1]{\accentset{\bullet}{#1}}
\newcommand{\st}[1]{\accentset{\times}{#1}}

\newcommand{\xPPN}{\emph{xPPN}}
\newcommand{\xAct}{\emph{xAct}}
\newcommand{\xTensor}{\emph{xTensor}}

\newcommand{\xPert}{\emph{xPert}}
\newcommand{\xPand}{\emph{xPand}}
\newcommand{\xTras}{\emph{xTras}}
\newcommand{\xCoba}{\emph{xCoba}}

\lstdefinelanguage[xPPN]{Mathematica}[]{Mathematica}{
morekeywords=[2]{OptionsPattern,OptionValue},
morekeywords=[3]{AllowUpperDerivatives,Antisymmetric,ChangeCovD,Christoffel,ConstantSymbolQ,ContractMetric,ContractMetrics,DefConstantSymbol,DefScalarFunction,DefTensor,delta,DependenciesOfTensor,IndicesOfVBundle,Labels,LI,LieDToCovD,MakeRule,NoScalar,OverDerivatives,ParamD,PD,PrintAs,Projected,ScalarFunctionQ,ScreenDollarIndices,SeparateMetric,SlotsOfTensor,Symmetric,SymmetryGroupOfTensor,ToCanonical,xTensorQ,Zero},
morekeywords=[4]{ApplyPPNRules,ApplyPPNRulesTo,Asym,BkgInvTetradM4,BkgInvTetradS3,BkgInvTetradT1,BkgMetricM4,BkgMetricS3,BkgMetricT1,BkgTetradM4,BkgTetradS3,BkgTetradT1,CD,ChristoffelCD,ChristoffelFD,ChristoffelND,Density,EnergyMomentum,FD,InternalEnergy,InvMet,InvTet,LorentzMfSpace,LorentzMfSpacetime,LorentzMfTime,Met,MetricToStandard,MfSpace,MfSpacetime,MfTime,ND,NonMet,OrderClear,OrderSet,OrderUnset,ParameterAlpha1,ParameterAlpha2,ParameterAlpha3,ParameterBeta,ParameterGamma,ParameterXi,ParameterZeta1,ParameterZeta2,ParameterZeta3,ParameterZeta4,PotentialA,PotentialB,PotentialChi,PotentialChiToPhiAB,PotentialChiToU,PotentialPhi1,PotentialPhi2,PotentialPhi3,PotentialPhi4,PotentialPhiW,PotentialToSource,PotentialU,PotentialUToChi,PotentialUToPhiAB,PotentialUToUU,PotentialUToV,PotentialUToW,PotentialUU,PotentialUUToChi,PotentialUUToU,PotentialV,PotentialVToChiW,PotentialVToU,PotentialVToW,PotentialW,PotentialWToChiV,PotentialWToU,PotentialWToV,PPN,PPNTensor,Pressure,RicciCD,SortPDs,SortPDsToBox,SortPDsToDiv,SortPDsToTime,SpaceTimeSplit,SpaceTimeSplits,TangentMfSpace,TangentMfSpacetime,TangentMfTime,Tet,TimePar,TimePiToEuler,TimeRhoToEuler,TimeVelToEuler,TREnergyMomentum,UsePPNRules,Velocity,VelocityOrder,Xi,T3a,T3b,T3c,T3d,T3z,L3A,L3Z}
}


\DeclareFontFamily{OT1}{pzc}{}
\DeclareFontShape{OT1}{pzc}{m}{it}{<-> s * [1.10] pzcmi7t}{}
\DeclareMathAlphabet{\mathpzc}{OT1}{pzc}{m}{it}

\begin{document}

\title{\xPPN: An implementation of the parametrized post-Newtonian formalism using \xAct{} for Mathematica}

\author{Manuel Hohmann}
\email{manuel.hohmann@ut.ee}
\affiliation{Laboratory of Theoretical Physics, Institute of Physics, University of Tartu, W. Ostwaldi 1, 50411 Tartu, Estonia}

\begin{abstract}
We present a package for the computer algebra system Mathematica, which implements the parametrized post-Newtonian (PPN) formalism. This package, named \xPPN, is built upon the widely used tensor algebra package suite \xAct, and in particular the package \xTensor{} therein. The main feature of \xPPN{} is to provide functions to perform a proper $3+1$ decomposition of tensors, as well as a perturbative expansion in so-called velocity orders, which are central tasks in the PPN formalism. Further, \xPPN{} implements various rules for quantities appearing in the PPN formalism, which aid in perturbatively solving the field equations of any metric theory of gravity. Besides Riemannian geometry, also teleparallel and symmetric teleparallel geometry are implemented.
\end{abstract}

\maketitle

\section{Introduction}\label{sec:intro}
The parametrized post-Newtonian (PPN) formalism~\cite{Nordtvedt:1968qs,Thorne:1970wv,Will:1971zzb,Will:1971wt,Will:1993ns,Will:2014kxa,Will:2018bme} is an indispensable tool for testing the viability of gravity theories. It is used to characterize any given theory of gravity by a set of ten parameters, which are closely related to the phenomenological properties of the theory under consideration. This allows to compare the parameters which are obtained theoretical through the PPN formalism with their values measured in experiments within the solar system and related physical settings.

In order for the PPN formalism to be applicable, the gravity theory under consideration must satisfy a number of assumptions. The most important is the existence of a metric governing the motion of test bodies, which can be described by a perturbation around a flat Minkowski background. Another assumption concerns the source of gravity, which is chosen to be a fluid satisfying the covariant conservation of energy-momentum, encoded in the Euler equations of fluid dynamics. Further, it is assumed that the field equations are of second derivative order, or can be brought into this form, so that their solution take a known standard form in terms of particular Poisson-like integrals over the source matter distribution.

In order to determine the post-Newtonian limit of a given gravity theory, one must perform a $3+1$ split of its field equations (which are, in general, tensor equations) into space and time components, and then perform a perturbative expansion around a fixed vacuum solution. Depending on the structure of the field equations, both tasks may be challenging to perform by hand, and so the use of computer algebra comes as a useful tool. Due to the common assumptions on which the PPN formalism is based, and the similar steps to be applied to different theories of gravity, it appears natural to implement these common tasks into a general computational tool, which can then be used to calculate the post-Newtonian limit for any given theory. A number of functions to achieve this has already been implemented in Maple~\cite{Puetzfeld:2006qq}. The aim of \xPPN~\cite{xppn} is to provide another implementation, based on a more extensive framework for performing tensor calculations.

A very powerful tensor algebra software is the \xTensor{} package, which is part of the \xAct{} suite of Mathematica packages~\cite{xact}. It comes with numerous functions to define and manipulate tensorial expressions, and includes concepts such as induced metrics on hypersurfaces orthogonal to a vector field, or component calculations in \xCoba, which can be used to achieve a $3+1$ split of tensorial expressions. The former is employed, for example, for cosmological perturbation theory in the \xPand{} package~\cite{Pitrou:2013hga}, in conjunction with the \xPert{} package~\cite{Brizuela:2008ra} for general tensor perturbations. It thus appears natural to follow a similar approach to implement also the PPN formalism in \xAct. However, the latter comes with a few peculiarities which obstruct this strategy. The main difficulty lies in the different treatment of space and time in the PPN formalism, such as assigning different perturbation orders to space and time components of tensors. Another, albeit smaller issue is the different convention for counting the perturbation orders in \xPert{} and the PPN formalism.

Even though it is possible to implement the PPN formalism also using the existing functionality mentioned above, it appears simpler to overcome the aforementioned difficulties by using a different approach both to the $3+1$ split and the perturbative expansion, without using the induced metric framework or the \xCoba{} and \xPert{} packages. This is the approach followed by \xPPN. The key idea is to complement every tensor defined on the spacetime manifold by a number of tensors on a purely spatial manifold, which additionally depend on an external time parameter, and which represent the separated time and space components of the original spacetime tensor. Also derivatives are split into spatial derivatives and derivatives with respect to the time parameter. This approach allows an essentially different treatment of perturbations for space and time components. The aim of this article is to present this approach, as well as the implementation of the PPN formalism which is based on this framework.

This article is structured as follows. We start with a brief review of the PPN formalism and its extensions implemented by \xPPN{} in section~\ref{sec:ppn}. The general concepts on which this implementation is based are explained in section~\ref{sec:concepts}. The main functionality of \xPPN{} is laid out in section~\ref{sec:objects}, where we display the geometric objects defined by \xPPN, and section~\ref{sec:functions}, where we explain the functions provided for the common operations on tensor expressions. A complete usage example is given in section~\ref{sec:example}, which shows how to calculate the PPN parameters of a simple scalar-tensor theory of gravity. A summary and outlook towards possible future extensions are given in section~\ref{sec:conclusion}. For further reference, we provide notes on the implementation and internal operation of \xPPN{} in appendix~\ref{app:implementation}.

In order to make code listings more readable, the following syntax highlighting is used. Mathematica keywords are typeset in {\color{DarkBlue}blue}, \xAct{} keywords are typeset in {\color{DarkGreen}green} and \xPPN{} keywords are typeset in {\color{DarkRed}red}. We use lowercase letters for coordinate indices and uppercase letters for Lorentz indices, where in both cases Greek indices run from \(0\) to \(3\) and belong to spacetime, while Latin indices run from \(1\) to \(3\) and belong to space only.

\section{Parametrized post-Newtonian formalism}\label{sec:ppn}
The aim of the \xPPN{} package is to provide an implementation of the parametrized post-Newtonian (PPN) formalism and several of its geometric extensions. Here we briefly summarize these theoretical foundations. We first discuss the standard PPN formalism and the perturbative expansion of the metric in section~\ref{ssec:ppncurv}. An extended formulation using a tetrad as the fundamental field instead of the metric is shown in section~\ref{ssec:ppntet}. We then display its application to teleparallel gravity in section~\ref{ssec:ppntors} and symmetric teleparallel gravity in section~\ref{ssec:ppnnonmet}.

\subsection{Standard formalism for Riemannian geometry}\label{ssec:ppncurv}
There exist different versions of the PPN formalism. Here we adhere to its form presented in~\cite{Will:1993ns}. Its starting point is the assumption that the propagation of light and massive test bodies is governed by a pseudo-Riemannian metric \(g_{\alpha\beta}\) of Lorentzian signature. This metric is further considered in a weak field limit as an asympotically flat perturbation around a flat Minkowski background. The source of this perturbation is assumed to be of compact support and modeled by the energy-momentum tensor
\begin{equation}\label{eq:fluid}
\Theta^{\alpha\beta} = (\rho + \rho\Pi + p)u^{\alpha}u^{\beta} + pg^{\alpha\beta}
\end{equation}
of a perfect fluid with four-velocity \(u^{\alpha}\), rest energy density \(\rho\), specific internal energy \(\Pi\) and pressure \(p\). Further, a fixed Cartesian coordinate system \((x^{\alpha}) = (t, x^a)\) is used, denoted as the \emph{universe rest frame}. It is then assumed that the velocity \(v^a = u^a/u^0\) of the source matter in this coordinate system is small compared to the speed of light, \(|v^a| \ll c \equiv 1\). This velocity takes the role of the perturbation parameter. Physical quantities, such as the metric, are expanded in terms of the source velocity, and any term in this perturbative expansion which is proportional to \(|v^a|^n\) is said to be of $n$'th velocity order, which is commonly denoted \(\mathcal{O}(n)\) in the literature\footnote{Note in particular that \(\mathcal{O}(1) \sim |v|\) in this notation is the first velocity order, and not unity, which would be \(\mathcal{O}(0) \sim 1\).}. For the metric \(g_{\alpha\beta}\), this expansion may be written explicitly in the form
\begin{equation}\label{eq:metvelord}
g_{\alpha\beta} = \sum_{n = 0}^{\infty}\order{g}{n}_{\alpha\beta} = \order{g}{0}_{\alpha\beta} + \order{g}{1}_{\alpha\beta} + \order{g}{2}_{\alpha\beta} + \order{g}{3}_{\alpha\beta} + \order{g}{4}_{\alpha\beta} + \mathcal{O}(5)\,,
\end{equation}
where we used overscripts to indicate velocity orders \(\order{g}{n}_{\alpha\beta} \sim \mathcal{O}(n)\). Terms beyond the fourth velocity order are usually not considered in the standard PPN formalism implemented by \xPPN. The zeroth velocity order is assumed to be the flat Minkowski background,
\begin{equation}\label{eq:metbkg}
\order{g}{0}_{\alpha\beta} = \eta_{\alpha\beta} = \mathrm{diag}(-1, 1, 1, 1)\,.
\end{equation}
For the metric perturbations, one finds that the first velocity order vanishes, \(\order{g}{1}_{\alpha\beta} = 0\), since the lowest velocity order of the matter source is the second order, as we will argue below. Further, the components \(\order{g}{2}_{0a}, \order{g}{3}_{00}, \order{g}{3}_{ab}, \order{g}{4}_{0a}\) change their sign under time reversal, and are prohibited by energy-momentum conservation. It follows that only the components
\begin{equation}\label{eq:metpert}
\order{g}{2}_{00}\,, \quad
\order{g}{2}_{ab}\,, \quad
\order{g}{3}_{0a}\,, \quad
\order{g}{4}_{00}\,, \quad
\order{g}{4}_{ab}
\end{equation}
are non-vanishing. For the first four terms, a particular standard gauge is assumed, in which the metric components take the form
\begin{subequations}\label{eq:standardppn}
\begin{align}
\order{g}{2}_{00} &= 2U\,,\label{eq:standardppn200}\\
\order{g}{2}_{ab} &= 2\gamma U\delta_{ab}\,,\label{eq:standardppn2ij}\\
\order{g}{3}_{0a} &= -\frac{1}{2}(3 + 4\gamma + \alpha_1 - \alpha_2 + \zeta_1 - 2\xi)V_a - \frac{1}{2}(1 + \alpha_2 - \zeta_1 + 2\xi)W_a\,,\label{eq:standardppn30i}\\
\order{g}{4}_{00} &= -2\beta U^2 + (2 + 2\gamma + \alpha_3 + \zeta_1 - 2\xi)\Phi_1 + 2(1 + 3\gamma - 2\beta + \zeta_2 + \xi)\Phi_2\nonumber\\
&\phantom{=}+ 2(1 + \zeta_3)\Phi_3 + 2(3\gamma + 3\zeta_4 - 2\xi)\Phi_4 - 2\xi\Phi_W - (\zeta_1 - 2\xi)\mathcal{A}\,.\label{eq:standardppn400}
\end{align}
\end{subequations}
For the last term \(\order{g}{4}_{ab}\) no such expansion is assumed in the standard PPN formalism, but it may be included in an extended formalism~\cite{Richter:1982zz,Richter:1982zza}. The terms on the right hand side are the so-called PPN potentials and PPN parameters; see sections~\ref{ssec:potdef} and~\ref{ssec:ppnpar} for their definition and~\cite{Will:1993ns} for a physical explanation. Here the PPN potentials describe the matter distribution, and their form is independent of the theory under consideration, while the PPN parameters are independent of the matter distribution, and their values are determined by the gravity theory. In order to determine the values of the PPN parameters for a given gravity theory, one follows a well-defined procedure to expand the gravitational field equations into velocity orders, which are then solved successively, up to the fourth order. The virtue of the PPN formalism is the fact that this form of the metric is sufficiently generic to solve the metric field equations of a large number of gravity theories, in which the source of gravity is the matter energy-momentum. In order to obtain this solution, one must also decompose the energy-momentum tensor~\eqref{eq:fluid} into space and time components, as well as into velocity orders. For the matter variables, one assigns the orders \(\rho \sim \Pi \sim \mathcal{O}(2)\) and \(p \sim \mathcal{O}(4)\), based on their order of magnitude in the solar system. Lowering the indices of the energy-momentum tensor, its components then take the form
\begin{subequations}\label{eq:energymomentum}
\begin{align}
\Theta_{00} &= \rho\left(1 + \Pi + |v|^2 - \order{g}{2}_{00}\right) + \mathcal{O}(6)\,,\\
\Theta_{0a} &= -\rho v_a + \mathcal{O}(5)\,,\\
\Theta_{ab} &= \rho v_av_b + p\delta_{ab} + \mathcal{O}(6)\,.
\end{align}
\end{subequations}
Further, one assumes that the gravitational field is quasi-static, which means that it changes only due to the motion of the source matter, excluding, e.g., transient gravitational waves. Hence, time derivatives of all physical quantities are weighted with an additional velocity order, \(\partial_0 \sim \mathcal{O}(1)\). In particular, this affects derivatives of the metric, which enter the field equations through the Levi-Civita connection
\begin{equation}\label{eq:levicivita}
\lc{\Gamma}^{\gamma}{}_{\alpha\beta} = \frac{1}{2}g^{\gamma\delta}(\partial_{\alpha}g_{\delta\beta} + \partial_{\beta}g_{\alpha\delta} - \partial_{\delta}g_{\alpha\beta})
\end{equation}
and its curvature tensor. Here and in the following, we denote terms related to the Levi-Civita connection with an empty circle, in order to distinguish them from other connections to be introduced later, following the standard convention in the literature on teleparallel and related gravity theories, where this distinction becomes relevant; note that this circle is a distinct symbol from a zero denoting the zeroth velocity order.

\subsection{Tetrad extension}\label{ssec:ppntet}
There exist numerous gravity theories in which one of the fundamental fields is a tetrad (or coframe) field \(\theta^{\Gamma}{}_{\alpha}\), which defines the metric via the relation
\begin{equation}\label{eq:metrictetrad}
g_{\alpha\beta} = \eta_{\Gamma\Delta}\theta^{\Gamma}{}_{\alpha}\theta^{\Delta}{}_{\beta}\,,
\end{equation}
where \(\eta_{\Gamma\Delta} = \mathrm{diag}(-1, 1, 1, 1)\) is again the Minkowski metric, here with Lorentz indices. Its post-Newtonian expansion can be obtained as follows~\cite{Nitsch:1979qn,Smalley:1980em,Hayward:1981bk}. Its zeroth order is given by a diagonal background tetrad,
\begin{equation}
\order{\theta}{0}^{\Gamma}{}_{\alpha} = \Delta^{\Gamma}{}_{\alpha} = \mathrm{diag}(1, 1, 1, 1)\,.
\end{equation}
defined in the previous section. For any higher order terms \(\order{\theta}{k}^{\Gamma}{}_{\alpha}\) in its perturbative expansion, it turns out to be more convenient to work with the expressions~\cite{Ualikhanova:2019ygl,Hohmann:2020vcv}
\begin{equation}
\order{\tau}{k}_{\alpha\beta} = \eta_{\alpha\gamma}\Delta_{\Gamma}{}^{\gamma}\order{\theta}{k}^{\Gamma}{}_{\beta}\,,
\end{equation}
which carry only spacetime indices, where
\begin{equation}
\Delta_{\Gamma}{}^{\alpha} = \mathrm{diag}(1, 1, 1, 1)
\end{equation}
is the inverse background tetrad. While it is entirely possible to use the perturbations \(\order{\tau}{k}_{\alpha\beta}\) as fundamental variables, it turns out to be more convenient to decompose them into metric perturbations and another, independent degree of freedom. This can be achieved by noting that the metric perturbations follow from the relation~\eqref{eq:metrictetrad} to be given by
\begin{equation}
\order{g}{n}_{\alpha\beta} = \eta_{\Gamma\Delta}\sum_{k = 0}^{n}\order{\theta}{k}^{\Gamma}{}_{\alpha}\order{\theta}{n-k}^{\Delta}{}_{\beta} = \eta^{\gamma\delta}\sum_{k = 0}^{n}\order{\tau}{k}_{\gamma\alpha}\order{\tau}{n-k}_{\delta\beta}\,.
\end{equation}
Hence, the $n$'th order metric perturbation encodes \(\order{g}{n}_{\alpha\beta}\) the symmetric part \(2\order{\tau}{n}_{(\alpha\beta)}\) of the tetrad perturbation, up to lower order terms. Isolating the highest orders from the sum on the right hand side, we find that this symmetric part is given by
\begin{equation}
\order{\tau}{n}_{\alpha\beta} + \order{\tau}{n}_{\beta\alpha} = \order{g}{n}_{\alpha\beta} - \eta^{\gamma\delta}\sum_{k = 1}^{n - 1}\order{\tau}{k}_{\gamma\alpha}\order{\tau}{n-k}_{\delta\beta}\,.
\end{equation}
For the antisymmetric part, which constitute the aforementioned independent degree of freedom, one may define another, antisymmetric tensor field by
\begin{equation}
\order{a}{n}_{\alpha\beta} = 2\order{\tau}{n}_{[\alpha\beta]} = \order{\tau}{n}_{\alpha\beta} - \order{\tau}{n}_{\beta\alpha}\,.
\end{equation}
In summary, the perturbative orders \(\order{\theta}{n}^{\Gamma}{}_{\alpha}\) of the tetrad for \(n \geq 1\) are then defined as
\begin{equation}\label{eq:tetpert}
\order{\theta}{n}^{\Gamma}{}_{\alpha} = \frac{1}{2}\eta^{\beta\gamma}\Delta^{\Gamma}{}_{\beta}\left(\order{g}{n}_{\gamma\alpha} + \order{a}{n}_{\gamma\alpha} - \eta_{\Delta\Theta}\sum_{k = 1}^{n - 1}\order{\theta}{k}^{\Delta}{}_{\gamma}\order{\theta}{n-k}^{\Theta}{}_{\alpha}\right)
\end{equation}
in terms of \(\order{g}{n}_{\alpha\beta}\) and \(\order{a}{n}_{\alpha\beta}\). Following a similar line of argument as for the metric, the only non-vanishing components of the antisymmetric tensor up to the fourth velocity order which must be considered are given by
\begin{equation}
\order{a}{2}_{ab}\,, \quad
\order{a}{3}_{0a}\,, \quad
\order{a}{4}_{ab}\,.
\end{equation}
Note finally that the tetrad \(\theta^{\Gamma}{}_{\alpha}\) comes also with an inverse, the frame field \(e_{\Gamma}{}^{\alpha}\). It follows from its relation with the coframe field that its background is the diagonal element \(\order{e}{0}_{\Gamma}{}^{\alpha} = \Delta_{\Gamma}{}^{\alpha}\), while higher perturbation orders are recursively defined as
\begin{equation}\label{eq:invtetpert}
\order{e}{n}_{\Gamma}{}^{\alpha} = -\Delta_{\Gamma}{}^{\beta}\sum_{k = 0}^{n - 1}\order{e}{k}_{\Delta}{}^{\alpha}\order{\theta}{n-k}^{\Delta}{}_{\beta}\,,
\end{equation}
where the tetrad perturbations on the right hand side are further expanded using the rule~\eqref{eq:tetpert}.

\subsection{Teleparallel geometry}\label{ssec:ppntors}
The tetrad and its perturbative expansion discussed above play a fundamental role as the dynamical field variable in a particular class of gravity theories known as teleparallel gravity~\cite{Aldrovandi:2013wha}. In their covariant formulation~\cite{Krssak:2018ywd}, another fundamental field variable is used, which is the spin connection \(\tp{\omega}^{\Gamma}{}_{\Delta\alpha}\). Together with the tetrad and its inverse, it defines another connection, whose coefficients are given by\footnote{Here we follow the convention used by \xTensor, in which the first lower index \(\alpha\) on the connection coefficients \(\tp{\Gamma}^{\gamma}{}_{\alpha\beta}\) is the ``derivative'' index.}
\begin{equation}\label{eq:teleaffconn}
\tp{\Gamma}^{\gamma}{}_{\alpha\beta} = e_{\Gamma}{}^{\gamma}\left(\partial_{\alpha}\theta^{\Gamma}{}_{\beta} + \tp{\omega}^{\Gamma}{}_{\Delta\alpha}\theta^{\Delta}{}_{\beta}\right)\,.
\end{equation}
The spin connection is further restricted to be flat and antisymmetric, so that also the affine connection has vanishing curvature and is metric-compatible,
\begin{equation}
\partial_{\alpha}\tp{\Gamma}^{\gamma}{}_{\beta\delta} - \partial_{\beta}\tp{\Gamma}^{\gamma}{}_{\alpha\delta} + \tp{\Gamma}^{\gamma}{}_{\alpha\lambda}\tp{\Gamma}^{\lambda}{}_{\beta\delta} - \tp{\Gamma}^{\gamma}{}_{\beta\lambda}\tp{\Gamma}^{\lambda}{}_{\alpha\delta} = 0\,, \quad
\tp{\nabla}_{\gamma}g_{\alpha\beta} = 0\,,
\end{equation}
but it possesses non-vanishing torsion. Under these conditions, the spin connection turns out to be a gauge quantity, so that one can locally choose a Lorentz gauge in which it vanishes, known as the Weitzenböck gauge~\cite{Golovnev:2017dox}. Choosing this gauge, \(\tp{\omega}^{\Gamma}{}_{\Delta\alpha} \equiv 0\), the connecting coefficients~\eqref{eq:teleaffconn} are expressed in terms of the tetrad and its inverse alone~\cite{Ualikhanova:2019ygl}. This allows to derive their perturbative expansion, and the perturbative expansion of any derived tensor fields, in terms of the perturbations~\eqref{eq:tetpert} and~\eqref{eq:invtetpert}, and hence in terms of \(\order{g}{n}_{\alpha\beta}\) and \(\order{a}{n}_{\alpha\beta}\).

\subsection{Symmetric teleparallel geometry}\label{ssec:ppnnonmet}
Finally, the third class of gravity theories discussed here has become known as symmetric teleparallel gravity theories~\cite{Nester:1998mp}. In these theories yet another connection is employed, whose coefficients we denote by \(\st{\Gamma}^{\gamma}{}_{\alpha\beta}\), and which is assumed to have vanishing curvature and torsion,
\begin{equation}
\partial_{\alpha}\st{\Gamma}^{\gamma}{}_{\beta\delta} - \partial_{\beta}\st{\Gamma}^{\gamma}{}_{\alpha\delta} + \st{\Gamma}^{\gamma}{}_{\alpha\lambda}\st{\Gamma}^{\lambda}{}_{\beta\delta} - \st{\Gamma}^{\gamma}{}_{\beta\lambda}\st{\Gamma}^{\lambda}{}_{\alpha\delta} = 0\,, \quad
\st{\Gamma}^{\gamma}{}_{\alpha\beta} - \st{\Gamma}^{\gamma}{}_{\beta\alpha} = 0\,.
\end{equation}
It follows from these properties that there exists a local coordinate system \((x'^{\alpha})\), called the \emph{coincident gauge}, such that its connection coefficients \(\st{\Gamma}'^{\gamma}{}_{\alpha\beta}\) in this coordinate system vanish~\cite{BeltranJimenez:2017tkd}. Hence, in the standard PPN coordinate system \((x^{\alpha})\), the connection coefficients are given by
\begin{equation}
\st{\Gamma}^{\gamma}{}_{\alpha\beta} = \frac{\partial x^{\gamma}}{\partial x'^{\delta}}\frac{\partial x'^{\delta}}{\partial x^{\alpha}\partial x^{\beta}}\,.
\end{equation}
The post-Newtonian expansion of these connection coefficients is derived from the assumption that the two coordinate systems \((x^{\alpha})\) and \((x'^{\alpha})\) are related by an infinitesimal coordinate transformation, induced by a vector field \(\xi^{\alpha}\), so that up to the quadratic order one can write
\begin{equation}
x'^{\alpha} = x^{\alpha} + \xi^{\alpha} + \frac{1}{2}\xi^{\beta}\partial_{\beta}\xi^{\alpha} + \ldots
\end{equation}
From this assumption follows that the connection coefficients take the form
\begin{equation}\label{eq:symgampert}
\st{\Gamma}^{\gamma}{}_{\alpha\beta} = \partial_{\alpha}\partial_{\beta}\xi^{\gamma} + \frac{1}{2}\left(\xi^{\delta}\partial_{\alpha}\partial_{\beta}\partial_{\delta}\xi^{\gamma} + 2\partial_{(\alpha}\xi^{\delta}\partial_{\beta)}\partial_{\delta}\xi^{\gamma} - \partial_{\alpha}\partial_{\beta}\xi^{\delta}\partial_{\delta}\xi^{\gamma}\right) + \ldots\,,
\end{equation}
also up to the quadratic order in the vector field \(\xi^{\alpha}\). Finally, the vector field is expanded in velocity orders. It turns out that in the PPN formalism the only non-vanishing components are given by~\cite{Flathmann:2020zyj}
\begin{equation}\label{eq:xipert}
\order{\xi}{2}^a\,, \quad
\order{\xi}{3}^0\,, \quad
\order{\xi}{4}^a\,.
\end{equation}
From these components the perturbative expansion of the connection coefficients \(\st{\Gamma}^{\gamma}{}_{\alpha\beta}\) is obtained.

\section{Mathematical foundation}\label{sec:concepts}
Two fundamental mathematical concepts form the basis of the PPN formalism detailed in the previous section, and so their implementation is an important part of any computational tool to address this formalism. Here we briefly discuss these concepts and their formulation we choose to implement in \xPPN. This will serve as an explanation of the mathematical background for the following sections. We discuss the split of tensors on spacetime into their space and time parts in section~\ref{ssec:31split} and the perturbative expansion into velocity order in section~\ref{ssec:perturb}.

\subsection{$3+1$ split of tensors}\label{ssec:31split}
As explained in the previous section, one of the crucial steps in applying the PPN formalism is the $3+1$ decomposition of all tensorial quantities. A proper, geometric interpretation of this split can be given by regarding the spacetime manifold \(M_4\), equipped with coordinates \((x^{\alpha}) = (t, x^a)\), as a product manifold \(M_4 \cong T_1 \times S_3\), where \(t\) is the time coordinate on the manifold \(T_1 \cong \mathbb{R}\), while \((x^a)\) are the spatial coordinates on \(S_3\). It follows that for every \(t \in T_1\), one can find a map \(i_t: S_3 \to M_4, (x^a) \mapsto (t, x^a)\), whose image in \(M_4\) is called the spatial slice, or constant time hypersurface at time \(t\). The collection of all spatial slices then forms a foliation of \(M_4\).

While the construction appears abstract, it gives a clear interpretation of the $3+1$ split of tensor fields as follows. A vector field with components \(X^{\alpha}\) on \(M_4\) splits in the form
\begin{equation}
X^{\alpha}\partial_{\alpha} = X^0\partial_0 + X^a\partial_a
\end{equation}
into a temporal part \(X^0\partial_0\), which is parallel to the coordinate lines generated by the time coordinate \(t\), and a spatial part \(X^a\partial_a\), which is tangent to the spatial hypersurfaces. Under purely spatial coordinate transformations, \(X^0\) behaves as a scalar, while \(X^a\) are the components of a vector field. For any fixed value of \(t\), we may thus take the pullback along \(i_t\) to obtain a scalar function \(i_t^*X^0\), as well as a vector field \(i_t^*(X^a\partial_a)\), both now being tensor fields on \(S_3\), hence functions of the spatial coordinates \((x^a)\). The dependence of the components \(X^{\alpha}\) on the time coordinate \(t\) is still present as a dependence on the choice of the spatial hypersurface, and \(t\) is regarded as a parameter which corresponds to this choice, instead of a coordinate.

The interpretation of time \(t\) as a parameter instead of a coordinate is, of course, purely mathematical, but it allows for a number of simplification when it comes to the implementation of the $3+1$ split in computer algebra. Most importantly for the PPN formalism, it allows to treat time derivatives \(\partial_0\) essentially differently from spatial derivatives \(\partial_a\). This is necessary, because in the PPN formalism time derivatives are weighted with an additional velocity order, which is not the case for spatial derivatives. Another advantage becomes apparent when considering tensors with symmetries. For example, considering an antisymmetric tensor field \(A_{\alpha\beta}\), the $3+1$ split can be written as
\begin{equation}
A_{\alpha\beta}\dd x^{\alpha} \otimes \dd x^{\beta} = A_{00}\dd t \otimes \dd t + A_{0a}\dd t \otimes \dd x^a + A_{a0}\dd x^a \otimes \dd t + A_{ab}\dd x^a \otimes \dd x^b\,,
\end{equation}
where \(A_{00} = 0\) and \(A_{0a} = -A_{a0}\). Hence, the independent components can be described by a single covector field \(A_{0a}\) and antisymmetric tensor field \(A_{ab}\) on \(S_3\). Thus, the use of tensor symmetries in conjunction with the $3+1$ split allows to immediately discard vanishing components, which helps to simplify the calculation.

The outlined interpretation of the $3+1$ split and the decomposition of tensor fields on spacetime \(M_4\) into tensor fields on space \(S_3\), with an additional parameter dependence, and corresponding counting of velocity orders, are a central ingredient of \xPPN. In section~\ref{ssec:stvelsel} we explain how to select certain components from the $3+1$ decomposition of a tensor field, while section~\ref{ssec:stsplit} shows how to decompose arbitrary tensorial expressions. The implementation of these operations is outlined in appendix~\ref{app:intrep}, which shows the internal representation of decomposed tensor fields, while in section~\ref{app:31algo} we schematically explain the decomposition algorithm and its treatment of tensor symmetries.

\subsection{Perturbative expansion in velocity orders}\label{ssec:perturb}
Another important ingredient of the PPN formalism is the perturbative expansion of tensor fields into velocity orders. Similarly to the expansion~\eqref{eq:metvelord} for the metric, also any other tensor field \(A\) is expanded as
\begin{equation}\label{eq:genvecord}
A = \sum_{n = 0}^{\infty}\order{A}{n}\,,
\end{equation}
where each term is of the corresponding velocity order \(\order{A}{n} \sim \mathcal{O}(n)\), and where we omit tensor indices for brevity. Note that this is different from the convention used, e.g., in the \xPert{} package~\cite{Brizuela:2008ra}, where the perturbative expansion takes the form
\begin{equation}
A = \sum_{n = 0}^{\infty}\frac{\varepsilon^n}{n!}\Delta^n[A]\,,
\end{equation}
and thus has the explicit form of a Taylor expansion in the perturbation parameter \(\varepsilon\). We therefore list a few formulas which follow from the perturbative expansion~\eqref{eq:genvecord}. Most notably, for a tensor \(A = A_1 \cdots A_N\) given as a product of \(N\) tensors \(A_1, \ldots, A_N\), one has the $n$'th order term given by
\begin{equation}\label{eq:velordprod}
\order{A}{n} = \sum_{k_1 + \ldots + k_N = n}\prod_{i = 1}^{N}\order{A_i}{k_i}\,,
\end{equation}
where the orders \(k_i\) of the factors \(A_i\) run over non-negative integers. Another important relation is the expansion of expressions \(F = f(A)\), where \(f\) is a scalar, or more general a tensorial function. In this case a Taylor expansion of the function \(f\) around the background \(\order{A}{0}\) is used,
\begin{equation}
f(A) = \sum_{k = 0}^{\infty}\frac{\left(A - \order{A}{0}\right)^k}{k!}f^{(k)}\left(\order{A}{0}\right)\,,
\end{equation}
where \(f^{(k)}\) denotes the $k$'th derivative, and then the product formula~\eqref{eq:velordprod} is applied. For a function of \(N\) arguments, this formula naturally generalizes to
\begin{equation}\label{eq:velordfunc}
f(A_1, \ldots, A_N) = \sum_{k_1 = 0}^{\infty} \cdots \sum_{k_N = 0}^{\infty}f^{(k_1, \ldots, k_N)}\left(\order{A}{0}_1, \ldots, \order{A}{0}_N\right)\prod_{i = 1}^{N}\frac{\left(A_i - \order{A}{0}_i\right)^{k_i}}{k_i!}\,.
\end{equation}
Finally, time derivatives are weighted with an additional velocity order, and so for \(A' = \partial_0A\) one has
\begin{equation}\label{eq:velordtimed}
\order{A}{n}' = \partial_0\order{A}{n - 1}\,,
\end{equation}
and analogously for higher derivative orders, where the left hand side is to be understood as the $n$'th order term in the expansion of \(A'\).

In \xPPN, the perturbative expansion of tensor fields into velocity orders, which takes into account the rules outlined above, is implemented in the function \lstinline/VelocityOrder/, which is explained in detail in section~\ref{ssec:veldecomp}. A few notes on the implementation of this function are given in appendix~\ref{app:velalgo}.

\section{Geometric objects defined by \xPPN}\label{sec:objects}
The PPN formalism relies on the existence of a number of objects, such as the spacetime manifold, the physical metric and its background value, as well as a particular set of potentials and constant parameters, which are necessary for its application to any gravity theory. For convenience, these objects are already defined by \xPPN{} when the package is loaded, so that the user must define only those additional geometric objects which are specific to the gravity theory under consideration. Here we give an overview of these pre-defined objects. The manifolds corresponding to the split of spacetime into space and time, as well as a number of bundles over these manifolds, are given in section~\ref{ssec:bundles}. To define tensors on these manifolds and write tensorial expressions, \xAct{} relies on indices; section~\ref{ssec:indices} lists the indices defined by \xPPN{} for the given bundles. The geometric objects constituting the background geometry are listed in section~\ref{ssec:background}, while section~\ref{ssec:dyngeo} gives a detailed account of the dynamical geometry and its perturbative expansion. The energy-momentum tensor and its expansion in fluid variables is shown in section~\ref{ssec:enmomdef}. Finally, the objects constituting the standard PPN metric are the PPN potentials detailed in section~\ref{ssec:potdef} and the PPN parameters listed in section~\ref{ssec:ppnpar}.

\subsection{Manifolds and bundles}\label{ssec:bundles}
Tensors in \xTensor{} are defined with respect to a manifold. In order to implement the $3+1$ decomposition discussed in section~\ref{ssec:31split}, \xPPN{} defines three manifolds:
\begin{enumerate}
\item
\lstinline/MfTime/ is the one-dimensional time manifold \(T_1\).
\item
\lstinline/MfSpace/ is the three-dimensional space manifold \(S_3\). All tensors for which the $3+1$ split into time and space components has been carried out are defined on this manifold, with an additional dependence on the time parameter \lstinline/TimePar/.
\item
\lstinline/MfSpacetime/ is the four-dimensional spacetime manifold \(M_4 \cong T_1 \times S_3\). It is defined as a product manifold, so that sums over tensor indices of \(M_4\) may automatically be decomposed into sums over \(T_1\) and \(S_3\). All geometric objects which appear in the theory under consideration should be defined on this manifold.
\end{enumerate}
Each of these manifolds is canonically equipped with a tangent bundle; \xTensor{} defines these automatically, and they are named \lstinline/TangentMfTime/, \lstinline/TangentMfSpace/ and \lstinline/TangentMfSpacetime/, respectively, and denoted by prefixing the manifold symbol with \(\mathbb{T}\). In addition, \xPPN{} defines another vector bundle over each of these three manifolds, whose rank is the same as the dimension of the manifold, and which is used to define quantities carrying Lorentz indices. These bundles are named \lstinline/LorentzMfTime/, \lstinline/LorentzMfSpace/ and \lstinline/LorentzMfSpacetime/, respectively, and denoted by prefixing the manifold symbol with \(\mathbb{L}\).

\subsection{Indices}\label{ssec:indices}
Indices play an important role in \xTensor, since they are used to denote tensor expressions and establish the relation between tensor slots and vector bundles. Often a large number of indices is necessary for longer tensor expressions, and each of them must be defined as a symbol (possibly with an appended number). \xPPN{} addresses this problem by pre-defining a large number of indices using symbols which are unlikely to collide with other notation. In particular, the following indices are defined:
\begin{enumerate}
\item
\lstinline/LI[0]/ is used by \xPPN{} to denote the time component of the $3+1$ decomposed forms of tensors, and is printed as 0. From the viewpoint of \xTensor, this is a label index, which is not associated with any vector bundle and has no implied meaning. It is used because $3+1$ decompositions are defined on \(S_3\), and so only spatial indices can be associated with the tangent bundle. Also it may appear an arbitrary number of times and in any position in any tensor expression.
\item
\lstinline/\[ScriptT]/
(printed as $\mathpzc{t}$) is the generic index on the tangent bundle \(\mathbb{T}T_1\). \xTensor{} will use this to automatically generate as many indices \(\mathpzc{t}_1, \mathpzc{t}_2, \ldots\) as needed as an intermediate step when performing a $3+1$ decomposition of indices, before replacing them with 0.
\item
\lstinline/\[ScriptCapitalT]/
(printed as $\mathpzc{T}$) is the generic index on the Lorentz bundle \(\mathbb{L}T_1\). It is used by \xTensor{} in the same way as \(\mathpzc{t}\) on \(\mathbb{T}T_1\).
\item
Lowercase Latin letters \(a, \ldots, z\) (character codes 97--122) are used to denote indices on \(\mathbb{T}S_3\). \xPPN{} automatically defines the symbols \lstinline/T3a/, \ldots, \lstinline/T3z/ to denote these indices, so that there is no need for the user to manually declare any indices, while at the same time avoiding possible name conflicts and leaving single letters free for other use. The prefix ``\texttt{T3}'' is omitted when the symbol is printed in output form, so that only the index letter itself appears in printed output.
\item
Uppercase Latin letters \(A, \ldots, Z\) (character codes 65--90) are used to denote indices on \(\mathbb{L}S_3\). As for the lowercase indices listed in the previous item, \xPPN{} defines symbols \lstinline/L3A/, \ldots, \lstinline/L3Z/ for these uppercase indices, and omits the prefix ``\texttt{L3}'' when printing the indices in output form.
\item
Lowercase Greek letters \(\alpha, \ldots, \omega\) (character codes 945--969, 977, 981, 982, 1008, 1009, 1013) are used to denote indices on \(\mathbb{T}M_4\). The corresponding symbols defined by \xPPN{} are denoted \lstinline/T4<a>/, \ldots, \lstinline/T4<w>/.
\item
Uppercase Greek letters \(\mathrm{A}, \ldots, \Omega\) (character codes 913--929, 931--937) are used to denote indices on \(\mathbb{L}M_4\). The corresponding symbols defined by \xPPN{} are denoted \lstinline/L4<A>/, \ldots, \lstinline/L4<W>/.
\end{enumerate}
To obtain a list of pre-defined indices for any of these bundles, the \xTensor{} command \lstinline/IndicesOfVBundle/ may be used. For example, to list all indices defined on \(\mathbb{T}M_4\), use:
\begin{lstlisting}
In[]:= IndicesOfVBundle[TangentMfSpacetime]
\end{lstlisting}
The Mathematica command \lstinline/Names/, together with a string pattern, may also be used to list these indices. This will give a similar result as the aforementioned command:
\begin{lstlisting}
In[]:= Names["T4" ~~ _]
\end{lstlisting}

\subsection{Background geometry}\label{ssec:background}
\xPPN{} automatically defines a number of geometric objects corresponding to the background geometry, around which the perturbative PPN expansion is performed. On each of the three manifolds \(T_1, S_3, M_4\) a metric, a tetrad and an inverse tetrad are defined. They are denoted as given in table~\ref{tab:bkgobj}. Each of these geometric objects is defined to be constant with respect to the partial derivative \lstinline/PD/ on its respective manifold, so that \lstinline/PD/ applied to any of these objects vanishes. Further, contractions of a tetrad and its inverse are carried out automatically, and yield the corresponding \lstinline/delta/ tensor.

\begin{table}[htb]
\centering
\begin{tabular}{l|c|c|c}
Symbol & Definition & Manifold & Indices\\\hline
\lstinline/BkgMetricM4/ & $\eta_{\alpha\beta} = \mathrm{diag}(-1, 1, 1, 1)$ & $M_4$ & $(-\mathbb{T}M_4, -\mathbb{T}M_4)$\\
\lstinline/BkgMetricS3/ & $\delta_{ab} = \eta_{ab}$ & $S_3$ & $(-\mathbb{T}S_3, -\mathbb{T}S_3)$\\
\lstinline/BkgMetricT1/ & $\eta_{00} = -1$ & $T_1$ & $(-\mathbb{T}T_1, -\mathbb{T}T_1)$\\
\lstinline/BkgTetradM4/ & $\Delta^{\Gamma}{}_{\alpha} = \mathrm{diag}(1, 1, 1, 1)$ & $M_4$ & $(\mathbb{L}M_4, -\mathbb{T}M_4)$\\
\lstinline/BkgTetradS3/ & $\Delta^A{}_a$ & $S_3$ & $(\mathbb{L}S_3, -\mathbb{T}S_3)$\\
\lstinline/BkgTetradT1/ & $\Delta^0{}_0$ & $T_1$ & $(\mathbb{L}T_1, -\mathbb{T}T_1)$\\
\lstinline/BkgInvTetradM4/ & $\Delta_{\Gamma}{}^{\alpha} = \mathrm{diag}(1, 1, 1, 1)$ & $M_4$ & $(-\mathbb{L}M_4, \mathbb{T}M_4)$\\
\lstinline/BkgInvTetradS3/ & $\Delta_A{}^a$ & $S_3$ & $(-\mathbb{L}S_3, \mathbb{T}S_3)$\\
\lstinline/BkgInvTetradT1/ & $\Delta_0{}^0$ & $T_1$ & $(-\mathbb{L}T_1, \mathbb{T}T_1)$\\
\end{tabular}
\caption{Background geometric objects defined by \xPPN.}
\label{tab:bkgobj}
\end{table}

NB! Note that each of the background metrics is defined as the first metric on its respective manifold. Hence, it is also used by \xTensor{} in order to raise and lower indices, such that for a vector defined as \(A^{\alpha}\) one has \(A_{\alpha}A^{\alpha} = A^{\alpha}A^{\beta}\eta_{\alpha\beta}\), which can easily be verified by using the \xTensor{} command \lstinline/SeparateMetric/:
\begin{lstlisting}
In[]:= DefTensor[A[T4<a>], {MfSpacetime}]

In[]:= ToCanonical[SeparateMetric[][A[-T4<a>] A[T4<a>]]]
Out[]= $A^{\alpha}A^{\beta}\eta_{\alpha\beta}$
\end{lstlisting}
In order to raise and lower indices with the physical metric \(g_{\alpha\beta}\), the metric (as defined in the following section) \emph{must} be written out explicitly.

\subsection{Dynamical geometry}\label{ssec:dyngeo}
The background geometry detailed above is defined independently of any gravity theory and matter configuration. This is contrasted with the dynamical geometry which we discuss next. These are the dynamical fields which are used to mediate the gravitational interaction, and which are related to the matter source by the gravitational field equations. The most important of these geometric objects is the metric, which we discuss in section~\ref{sssec:metric}. A similar role may be taken by the tetrad, as shown in section~\ref{sssec:tetrad}. Further, three covariant derivatives are defined, which represent the three connections used in different formulations of general relativity~\cite{BeltranJimenez:2019tjy}: the Levi-Civita connection in section~\ref{sssec:levicivita}, the teleparallel connection in section~\ref{sssec:telep} and the symmetric teleparallel connection in section~\ref{sssec:symtelep}.

\subsubsection{Metric}\label{sssec:metric}
As detailed in section~\ref{ssec:ppncurv}, the central object in the PPN formalism is the metric \(g_{\alpha\beta}\), which is denoted by \lstinline/Met[-T4<a>, -T4<b>]/ in \xPPN. Its inverse \(g^{\alpha\beta}\) is written as \lstinline/InvMet[T4<a>, T4<b>]/. Also here it must be emphasized that this is different from \lstinline/Met[T4<a>, T4<b>]/, which would be interpreted differently:
\begin{lstlisting}
In[]:= ToCanonical[SeparateMetric[][Met[T4<a>, T4<b>]]]
Out[]= $\eta^{\alpha\gamma}\eta^{\beta\delta}g_{\gamma\delta}$
\end{lstlisting}
A number of rules are defined for the perturbative expansion of the metric, and automatically applied when this expansion is performed. The zeroth order reduces to the background Minkowski metric~\eqref{eq:metbkg}, and so its space and time components are given by
\begin{equation}
\order{g}{0}_{00} = -1\,, \quad
\order{g}{0}_{0a} = 0\,, \quad
\order{g}{0}_{ab} = \delta_{ab}\,.
\end{equation}
The remaining components, up to the fourth velocity order, are set to vanish, except for the components~\eqref{eq:metpert}, for which no further rules are defined. These must be solved for in order to determine the PPN parameters. See section~\ref{ssec:orderrules} for more information how to apply these rules using the function \lstinline/ApplyPPNRules/.

\subsubsection{Tetrad}\label{sssec:tetrad}
In order to implement the tetrad extension to the PPN formalism detailed in section~\ref{ssec:ppntet}, \xPPN{} defines the tetrad \(\theta^{\Gamma}{}_{\alpha}\) as the object \lstinline/Tet[L4<G>, -T4<a>]/, together with an appropriate set of rules to evaluate its perturbative expansion using the formula~\eqref{eq:tetpert}. The antisymmetric tensor field \(a_{\alpha\beta}\) occurring in this expansion is denoted by \lstinline/Asym[-T4<a>, -T4<b>]/ in \xPPN. Further, next to the tetrad, also its inverse \(e_{\Gamma}{}^{\alpha}\) is defined in \xPPN, which is entered as \lstinline/InvTet[-L4<G>, T4<a>]/. A number of automatically applied rules are defined for these inverse tetrads, so that they are contracted with tetrads if possible, and derivatives are evaluated according to
\begin{equation}
e_{\Gamma}{}^{\alpha}\theta^{\Gamma}{}_{\beta} = \delta^{\alpha}_{\beta}\,, \quad
e_{\Gamma}{}^{\alpha}\theta^{\Delta}{}_{\alpha} = \delta_{\Gamma}^{\Delta}\,, \quad
\partial_{\beta}e_{\Gamma}{}^{\alpha} = -e_{\Gamma}{}^{\gamma}e_{\Delta}{}^{\alpha}\partial_{\beta}\theta^{\Delta}{}_{\gamma}\,.
\end{equation}
Further, perturbations of the tetrad are evaluated following the formula~\eqref{eq:invtetpert}, together with expanding the tetrad perturbation using the expansion~\eqref{eq:tetpert}.

\subsubsection{Levi-Civita connection}\label{sssec:levicivita}
Together with the metric \lstinline/Met[-T4<a>, -T4<b>]/, \xPPN{} defines its unique metric-compatible, torsion-free Levi-Civita connection, whose coefficients are defined from the metric by the well-known formula~\eqref{eq:levicivita}. The corresponding covariant derivative is entered as \lstinline/CD[-T4<a>]/, and it is written in postfix notation using a semicolon ``;'', as well as using the symbol \(\lc{\nabla}\) in prefix notation, to distinguish it from other covariant derivatives introduced later. \xTensor{} automatically defines a number of tensors for any covariant derivative. Most notably are the Christoffel ``tensors'' \lstinline/ChristoffelCD[T4<g>, -T4<a>, -T4<b>]/, which are defined as the difference between the connection coefficients \(\lc{\Gamma}^{\gamma}{}_{\alpha\beta}\) and the (vanishing) coefficients of a fiducial connection associated to the partial derivatives \(\partial_{\alpha}\) with respect to the coordinates. Further, \xTensor{} defines a number of curvature tensors, such as the Riemann, Ricci, Einstein and Weyl tensors. \xPPN{} defines the corresponding perturbative expansions in terms of the metric perturbations for all tensor fields derived from the covariant derivative \lstinline/CD[-T4<a>]/.

\subsubsection{Teleparallel connection}\label{sssec:telep}
In order to work with teleparallel gravity theories, as discussed in section~\ref{ssec:ppntors}, \xPPN{} defines also the teleparallel connection~\eqref{eq:teleaffconn}, along with its torsion tensor and their perturbative expansion. The corresponding covariant derivative is entered as \lstinline/FD[-T4<a>]/, and is denoted \(\tp{\nabla}\) in prefix notation and a bar ``|'' in postfix notation. Working in the Weitzenböck gauge implies that the connection coefficients \lstinline/ChristoffelFD[T4<g>, -T4<a>, -T4<b>]/ take the simple form
\begin{equation}
\tp{\Gamma}^{\gamma}{}_{\alpha\beta} = e_{\Gamma}{}^{\gamma}\partial_{\alpha}\theta^{\Gamma}{}_{\beta}\,,
\end{equation}
and their perturbative expansion is defined accordingly.

\subsubsection{Symmetric teleparallel connection}\label{sssec:symtelep}
Finally, to accommodate the PPN formalism for symmetric teleparallel theories of gravity shown in section~\ref{ssec:ppnnonmet}, \xPPN{} provides yet another pre-defined connection, which is characterized by vanishing torsion and curvature, but which is not compatible with the metric \(g_{\mu\nu}\). Its covariant derivative is entered as \lstinline/ND[-T4<a>]/ and denoted by the symbol \(\st{\nabla}\) in prefix notation, and a hash ``\#'' in postfix notation. The perturbative expansion~\eqref{eq:symgampert} of its connection coefficients is expressed in terms of a vector field \(\xi^{\alpha}\), which is defined under the name \lstinline/Xi[T4<a>]/ in \xPPN. Its perturbative expansion is implemented such that the only non-vanishing components are the three components~\eqref{eq:xipert}. Also for this connection, \xAct{} automatically defines torsion and curvature tensors, but both are set to vanish identically. The only tensorial quantity describing this connection and its relation to the Levi-Civita connection is the nonmetricity, which is given by
\begin{equation}
Q_{\alpha\beta\gamma} = \st{\nabla}_{\alpha}g_{\beta\gamma}\,.
\end{equation}
In \xPPN, the nonmetricity is called \lstinline/NonMet[-T4<a>, -T4<b>, -T4<g>]/, and its perturbative expansion is obtained from the definition above.

\subsection{Energy-momentum variables}\label{ssec:enmomdef}
In the PPN formalism it is assumed that the source of gravity is the energy-momentum tensor \(\Theta_{\alpha\beta}\) of a perfect fluid~\eqref{eq:fluid}, which defined by \xPPN{} is a tensor on \(M_4\) denoted by \lstinline/EnergyMomentum[-T4<a>, -T4<b>]/. Note that it is defined with lower indices. In order to raise the indices, the physical metric \(g_{\alpha\beta}\) \emph{must} be used explicitly; implicitly raising the indices by writing \lstinline/EnergyMomentum[T4<a>, T4<b>]/ would use the background metric \(\eta_{\alpha\beta}\), and hence \emph{not} give the correct result. Also note that we use the symbol \(\Theta\) instead of the more conventional \(T\) in order to avoid confusion with the torsion tensor.

For convenience, \xPPN{} also defines the trace-reversed energy-momentum tensor
\begin{equation}
\bar{\Theta}_{\alpha\beta} = \Theta_{\alpha\beta} - \frac{1}{2}g_{\alpha\beta}g^{\gamma\delta}\Theta_{\gamma\delta}\,,
\end{equation}
denoted by \lstinline/TREnergyMomentum[-T4<a>, -T4<b>]/, and which is likewise defined as a tensor on \(M_4\).

In order to describe the $3+1$ split of the energy-momentum tensor and its decomposition into velocity orders, \xPPN{} further defines the following tensors on \(S_3\) depending on \lstinline/TimePar/, which represent the variables describing the perfect fluid:
\begin{enumerate}
\item
\lstinline/Density[]/ is the rest mass density \(\rho \sim \mathcal{O}(2)\).
\item
\lstinline/Pressure[]/ is the pressure \(p \sim \mathcal{O}(4)\).
\item
\lstinline/InternalEnergy[]/ is the specific internal energy \(\Pi \sim \mathcal{O}(2)\).
\item
\lstinline/Velocity[T3a]/ is the velocity \(v^a \sim \mathcal{O}(1)\).
\end{enumerate}
When the energy-momentum tensor \(\Theta_{\alpha\beta}\) is expanded in velocity orders as shown in section~\ref{ssec:veldecomp}, the relations
\begin{equation}
\order{\Theta}{2}_{00} = \rho\,, \quad
\order{\Theta}{4}_{00} = \rho\left(\Pi + |v|^2 - \order{g}{2}_{00}\right)\,, \quad
\order{\Theta}{3}_{0a} = -\rho v_a\,, \quad
\order{\Theta}{4}_{ab} = \rho v_av_b + p\delta_{ab}\,,
\end{equation}
which follow directly from the expansion~\eqref{eq:energymomentum}, are applied, while all other components vanish.

\subsection{Post-Newtonian potentials}\label{ssec:potdef}
The post-Newtonian potentials are a central ingredient to the PPN formalism. In \xPPN{} they are defined as tensors on the space manifold \(S_3\) with an additional time dependence. Most of them are scalars, but there are also vector and tensor potentials, which therefore carry indices in the tangent bundle \(\mathbb{T}S_3\). In particular, the following PPN potentials are defined:
\begin{enumerate}
\item
\lstinline/PotentialChi[]/ is the post-Newtonian superpotential
\begin{equation}
\chi(t, \vec{x}) = -\int \dd^3x'\rho(t, \vec{x}')|\vec{x} - \vec{x}'|\,.
\end{equation}
\item
\lstinline/PotentialU[]/ is the Newtonian potential
\begin{equation}
U(t, \vec{x}) = \int \dd^3x'\frac{\rho(t, \vec{x}')}{|\vec{x} - \vec{x}'|}\,.
\end{equation}
\item
\lstinline/PotentialUU[-T3a, -T3b]/ is the anisotropic potential
\begin{equation}
U_{ab}(t, \vec{x}) = \int \dd^3x'\frac{\rho(t, \vec{x}')}{|\vec{x} - \vec{x}'|^3}(x_a - x_a')(x_b - x_b') = \chi_{,ab} - \frac{1}{2}\triangle\chi\delta_{ab}\,.
\end{equation}
\item
\lstinline/PotentialV[-T3a]/ is the isotropic vector potential
\begin{equation}
V_a(t, \vec{x}) = \int \dd^3x'\frac{\rho(t, \vec{x}')v_a(t, \vec{x}')}{|\vec{x} - \vec{x}'|}\,.
\end{equation}
\item
\lstinline/PotentialW[-T3a]/ is the anisotropic vector potential
\begin{equation}
W_a(t, \vec{x}) = \int \dd^3x'\frac{\rho(t, \vec{x}')v_b(t, \vec{x}')(x_a - x_a')(x_b - x_b')}{|\vec{x} - \vec{x}'|^3}\,.
\end{equation}
\item
\lstinline/PotentialPhi1[]/ is the kinetic energy potential
\begin{equation}
\Phi_1(t, \vec{x}) = \int \dd^3x'\frac{\rho(t, \vec{x}')v(t, \vec{x}')^2}{|\vec{x} - \vec{x}'|}\,.
\end{equation}
\item
\lstinline/PotentialPhi2[]/ is the gravitational self-energy potential
\begin{equation}
\Phi_2(t, \vec{x}) = \int \dd^3x'\frac{\rho(t, \vec{x}')U(t, \vec{x}')}{|\vec{x} - \vec{x}'|}\,.
\end{equation}
\item
\lstinline/PotentialPhi3[]/ is the internal energy potential
\begin{equation}
\Phi_3(t, \vec{x}) = \int \dd^3x'\frac{\rho(t, \vec{x}')\Pi(t, \vec{x}')}{|\vec{x} - \vec{x}'|}\,.
\end{equation}
\item
\lstinline/PotentialPhi4[]/ is the pressure potential
\begin{equation}
\Phi_4(t, \vec{x}) = \int \dd^3x'\frac{p(t, \vec{x}')}{|\vec{x} - \vec{x}'|}\,.
\end{equation}
\item
\lstinline/PotentialA[]/ is the anisotropic kinetic potential
\begin{equation}
\mathcal{A}(t, \vec{x}) = \int \dd^3x'\frac{\rho(t, \vec{x}')\left[v_a(t, \vec{x}')(x_a - x_a')\right]^2}{|\vec{x} - \vec{x}'|^3}\,.
\end{equation}
\item
\lstinline/PotentialB[]/ is the potential related to change in velocity
\begin{equation}
\mathcal{B}(t, \vec{x}) = \int \dd^3x'\frac{\rho(t, \vec{x}')}{|\vec{x} - \vec{x}'|}(x_a - x_a')\frac{\dd v_a(t, \vec{x}')}{\dd t}\,.
\end{equation}
\item
\lstinline/PotentialPhiW[]/ is the Whitehead potential
\begin{equation}
\Phi_W(t, \vec{x}) = \int \dd^3x'\int \dd^3x''\rho(t, \vec{x}')\rho(t, \vec{x}'')\frac{x_a - x_a'}{|\vec{x} - \vec{x}'|^3}\left(\frac{x_a' - x_a''}{|\vec{x} - \vec{x}''|} - \frac{x_a - x_a''}{|\vec{x}' - \vec{x}''|}\right)\,.
\end{equation}
\end{enumerate}
Note that these integrals are not implemented directly in \xPPN, which makes no explicit use of the coordinates apart from the assumption that partial derivatives of the tensors defining the background geometry vanish. However, they enter into the formalism indirectly, since they define the relations between the potentials and the source terms which are explained in detail in section~\ref{ssec:pottrans}.

\subsection{Post-Newtonian parameters}\label{ssec:ppnpar}
The PPN parameters are represented in \xPPN{} by objects which are declared as constants within \xTensor, so that any derivatives acting on them vanish. A full list is given in table~\ref{tab:ppnpar}. Note that assuming the PPN parameters to be constant, which is one of the standard assumptions of the PPN formalism, means that it cannot be directly applied to, e.g., theories with massive fields mediating the gravitational interaction, since in such theories the values of the PPN parameters depend on the distance between the source and the test mass; also time variation of PPN parameters depending on a cosmological background is not considered. The former may be included in a future version of \xPPN{} by introducing additional Yukawa-type potentials, which depend on a mass determining the interaction scale~\cite{Zaglauer:1990yh,Helbig:1991pk}. The latter might be implemented by allowing for PPN parameters which depend on \lstinline/TimePar/ instead of being declared constant~\cite{Sanghai:2016tbi}.

\begin{table}[htb]
\centering
\begin{tabular}{l|c}
Symbol & Parameter\\\hline
\lstinline/ParameterBeta/ & $\beta$\\
\lstinline/ParameterGamma/ & $\gamma$\\
\lstinline/ParameterAlpha1/ & $\alpha_1$\\
\lstinline/ParameterAlpha2/ & $\alpha_2$\\
\lstinline/ParameterAlpha3/ & $\alpha_3$\\
\lstinline/ParameterZeta1/ & $\zeta_1$\\
\lstinline/ParameterZeta2/ & $\zeta_2$\\
\lstinline/ParameterZeta3/ & $\zeta_3$\\
\lstinline/ParameterZeta4/ & $\zeta_4$\\
\lstinline/ParameterXi/ & $\xi$
\end{tabular}
\caption{Constants representing the PPN parameters in \xPPN.}
\label{tab:ppnpar}
\end{table}

\section{Utility functions}\label{sec:functions}
We now present a number of utility functions defined by \xPPN, which can be used to manipulate terms which typically appear in the post-Newtonian expansion, perform necessary computational steps and solve the gravitational field equations in terms of the post-Newtonian potentials and parameters listed in the previous sections. Which of these functions need to be used highly depends on the gravity theory under consideration, and we show this for a few of them in an explicit example in section~\ref{sec:example}; here we give an overview of the functions provided and how they are applied. In section~\ref{ssec:stvelsel} we show how to refer to specific parts of tensors in their space-time decomposition and perturbative expansion. The definition and application of substitution rules for such terms is explained in section~\ref{ssec:orderrules}. The decomposition of general tensorial expressions into space and time components is shown in section~\ref{ssec:stsplit}, and further into velocity orders in section~\ref{ssec:veldecomp}. The transformation of terms using the Euler equations derived from energy-momentum conservation is displayed in section~\ref{ssec:euler}, and applied to post-Newtonian potentials in section~\ref{ssec:pottrans}. Finally, utility functions for sorting derivatives prior to applying these rules are shown in section~\ref{ssec:derivsort}.

\subsection{Selecting space-time components and velocity orders of tensors}\label{ssec:stvelsel}
As explained in section~\ref{ssec:31split}, \xPPN{} defines for each tensor declared on the spacetime manifold \(M_4\) a number of tensors on the space manifold \(S_3\), which represent the $3+1$ decomposition of the initial tensor and its velocity orders. In order to access these components, \xPPN{} defines the utility function \lstinline/PPN/ to easily obtain them from a tensor head, a sequence of indices and optionally a velocity order. This function can be used in two ways, taking the following arguments:
\begin{enumerate}
\item
\lstinline/PPN[$h$][$i$]/, where \(h\) is a tensor head and \(i\) is a sequence of indices, such that each index either belongs to \(S_3\) or is \lstinline/LI[0]/ (possible with a minus sign);
\item
\lstinline/PPN[$h$, $n$][$i$]/, where \(n\) is a non-negative integer and \(h\) and \(i\) are as above.
\end{enumerate}
The application of this function is illustrated by the following example of a vector \(A^{\alpha}\):
\begin{lstlisting}
In[]:= DefTensor[A[T4<a>], MfSpacetime]

In[]:= PPN[A][T3a]
Out[]= $A^a$

In[]:= PPN[A, 2][LI[0]]
Out[]= $\order{A}{2}^0$
\end{lstlisting}
Note that the indices do not have to be in the natural position in which they have been for the tensor head \(h\). The function \lstinline/PPN/ yields the same result as if the indices had been specified in their natural position, and then raised or lowed with the background metric \lstinline/BkgMetricS3/ on \(S_3\):
\begin{lstlisting}
In[]:= DefTensor[A[T4<a>], MfSpacetime]

In[]:= ContractMetric[PPN[A][T3b] BkgMetricS3[-T3b, -T3a]]
Out[]= $A_a$

In[]:= % == PPN[A][-T3a]
Out[]= True
\end{lstlisting}
Tensor symmetries are taken into account, and indices are sorted into canonical order if possible:
\begin{lstlisting}
In[]:= DefTensor[A[-T4<a>, -T4<b>], MfSpacetime, Antisymmetric[{1, 2}]]

In[]:= PPN[A][-LI[0], -LI[0]]
Out[]= $0$

In[]:= PPN[A][-T3a, -LI[0]]
Out[]= $-A_{0a}$
\end{lstlisting}

\subsection{Definition and application of replacement rules}\label{ssec:orderrules}
For pre-defined tensors on the spacetime manifold \(M_4\), such as the metric or the energy-momentum tensor, \xPPN{} defines a number of substitution rules for the terms in their post-Newtonian expansion. In order to apply these rules to any given tensorial expression, \xPPN{} provides the function \lstinline/ApplyPPNRules/, which can be invoked in two different ways:
\begin{enumerate}
\item
\lstinline/ApplyPPNRules[$X$]/ recursively applies the defined PPN rules to all tensors which appear in the expression \(X\) and its subexpressions.
\item
\lstinline/ApplyPPNRules[$X$, $h$]/ applies the defined PPN rules only to those tensors within \(X\) with head \(h\).
\end{enumerate}
For example, the zeroth order of the physical metric is given by the Minkowski background metric:
\begin{lstlisting}
In[]:= {PPN[Met, 0][-LI[0], -LI[0]], PPN[Met, 0][-LI[0], -T3a], PPN[Met, 0][-T3a, -T3b]}
Out[]= {$\order{g}{0}_{00}$, $\order{g}{0}_{0a}$, $\order{g}{0}_{ab}$}

In[]:= ApplyPPNRules /@ %
Out[]= {$-1$, $0$, $\delta_{ab}$}
\end{lstlisting}
By default, \xPPN does not associate any rules to the terms in the post-Newtonian expansion of user-defined tensors. Therefore, \xPPN{} provides a number of functions to define and undefine such rules. This can be done with the following functions:
\begin{enumerate}
\item
\lstinline/OrderSet[PPN[$h$, $n$][$i$], $X$]/ defines or replaces a rule, such that \lstinline/ApplyPPNRules/ substitutes the $n$'th order term \lstinline/PPN[$h$, $n$][$i$]/ by $X$, where $X$ can be any tensorial expression which has the same free indices as defined by \(i\).
\item
\lstinline/OrderUnset[PPN[$h$, $n$][$i$]]/ removes the PPN rule associated with the term \lstinline/PPN[$h$, $n$][$i$]/ in the post-Newtonian expansion.
\item
\lstinline/OrderClear[$h$]/ removes any PPN rules associated to the tensor head \(h\).
\end{enumerate}
Their application can be illustrated as follows.
\begin{lstlisting}
In[]:= DefTensor[A[T4<a>], MfSpacetime]
In[]:= a = {PPN[A, 0][LI[0]], PPN[A, 0][T3a], PPN[A, 1][LI[0]], PPN[A, 1][T3a]};
In[]:= ApplyPPNRules /@ a
Out[]= {$\order{A}{0}^0$, $\order{A}{0}^a$, $\order{A}{1}^0$, $\order{A}{1}^a$}

In[]:= OrderSet[PPN[A, 0][LI[0]], 1];
In[]:= OrderSet[PPN[A, 0][T3a], 0];
In[]:= OrderSet[PPN[A, 1][T3a], Velocity[T3a]];
In[]:= ApplyPPNRules /@ a
Out[]= {$1$, $0$, $\order{A}{1}^0$, $v^a$}

In[]:= OrderUnset[PPN[A, 1][T3a]];
In[]:= ApplyPPNRules /@ a
Out[]= {$1$, $0$, $\order{A}{1}^0$, $\order{A}{1}^a$}

In[]:= OrderClear[A];
In[]:= ApplyPPNRules /@ a
Out[]= {$\order{A}{0}^0$, $\order{A}{0}^a$, $\order{A}{1}^0$, $\order{A}{1}^a$}
\end{lstlisting}

\subsection{$3+1$ space-time split of tensorial expressions}\label{ssec:stsplit}
\xPPN{} defines two functions handling the $3+1$ decomposition of arbitrary tensorial expressions on spacetime into their space and time components, as explained in section~\ref{ssec:31split}: \lstinline/SpaceTimeSplit/ and \lstinline/SpaceTimeSplits/. While the former calculates one specific component of the $3+1$ decomposition, the latter yields an array with all components. In both functions the decomposition is applied to both free and dummy indices.

The function \lstinline/SpaceTimeSplit/ takes two arguments: a tensor on \(M_4\) (possibly containing free indices) and a list of replacement rules, which assigns to each free index in a bundle over \(M_4\) either an index over the corresponding bundle over \(S_3\) or the label index \lstinline/LI[0]/ (possibly dressed with a minus sign to indicate a lower index). For example, for an expression of the form \(A^{\alpha}{}_{\beta}\) with free indices \lstinline/T4<a>/, \lstinline/-T4<b>/, the second argument may be any of the following:
\begin{enumerate}
\item \lstinline/{T4<a> -> LI[0], -T4<b> -> -LI[0]}/
\item \lstinline/{T4<a> -> T3a, -T4<b> -> -LI[0]}/
\item \lstinline/{T4<a> -> LI[0], -T4<b> -> -T3b}/
\item \lstinline/{T4<a> -> T3a, -T4<b> -> -T3b}/
\end{enumerate}
Of course, other names than \(a, b\) may be used for the indices on \(S_3\), but their position must remain the same; their role is to specify how the free indices in the resulting expression will be named. The function \lstinline/SpaceTimeSplit/ then calculates the component of the expression where each free index on \(M_4\) is replaced by the specified indices of the $3+1$ decomposition. For example, for a tensor \(A^{\alpha}{}_{\beta}\) defined by
\begin{lstlisting}
In[]:= DefTensor[A[T4<a>, -T4<b>], MfSpacetime]
\end{lstlisting}
one may use
\begin{lstlisting}
In[]:= SpaceTimeSplit[A[T4<a>, -T4<b>], {T4<a> -> T3a, -T4<b> -> -LI[0]}]
Out[]= $A^a{}_0$

In[]:= % == PPN[A][T3a, -LI[0]]
Out[]= True
\end{lstlisting}
to obtain the component $A^a{}_0$. In contrast to the function \lstinline/PPN/, which yields the $3+1$ decomposed component of a single tensor only, any tensorial expression may be decomposed by \lstinline/SpaceTimeSplit/ instead of a single tensor. For example, to decompose the expression \(A^{\alpha}{}_{\gamma}A^{\gamma}{}_{\beta}\) one may use
\begin{lstlisting}
In[]:= SpaceTimeSplit[A[T4<a>, -T4<g>] A[T4<g>, -T4<b>], {T4<a> -> T3a, -T4<b> -> -LI[0]}]
Out[]= $A^a{}_0A^0{}_0 + A^a{}_bA^b{}_0$
\end{lstlisting}
Observe that also the dummy index \(\gamma\) has been split into a sum over dummy indices in the $3+1$ decomposition.

The function \lstinline/SpaceTimeSplits/ is similar, but in its second argument \emph{only} indices of \(S_3\) (and hence no \lstinline/LI[0]/) are allowed as the right hand sides of the rules. The function then returns an array in which the free indices of the original expression are replaced either by the specified spatial indices or the time index 0, as shown in the following example:
\begin{lstlisting}
In[]:= SpaceTimeSplits[A[T4<a>, -T4<b>], {T4<a> -> T3a, -T4<b> -> -T3b}]
Out[]= {{$A^0{}_0$, $A^0{}_b$}, {$A^a{}_0$, $A^a{}_b$}}
\end{lstlisting}
The order of the rules in the replacement list corresponds to the order of the dimensions of the resulting array. In the example above, the first dimension corresponds to \(\alpha\), while the second dimension corresponds to \(\beta\). Hence,
\begin{lstlisting}
In[]:= %[[1, 2]]
Out[]= $A^0{}_b$
\end{lstlisting}
selects the component where \(\alpha\) is replaced by \(0\) (the first element of \((0, a)\)), while \(\beta\) is replaced by \(b\) (the second element of \((0, b)\)). Similarly,
\begin{lstlisting}
In[]:= SpaceTimeSplits[A[T4<a>, -T4<b>], {-T4<b> -> -T3b, T4<a> -> T3a}]
Out[]= {{$A^0{}_0$, $A^a{}_0$}, {$A^0{}_b$, $A^a{}_b$}}
\end{lstlisting}
yields the transposed array.

Finally, we remark that both \lstinline/SpaceTimeSplit/ and \lstinline/SpaceTimeSplits/ also operate on partial derivatives \lstinline/PD[-T4<a>]/ with respect to spacetime coordinates. These are converted as follows:
\begin{lstlisting}
In[]:= DefTensor[A[], MfSpacetime]
In[]:= PD[-T4<a>][A[]]
Out[]= $\partial_{\alpha}A$

In[]:= SpaceTimeSplits[%, {-T4<a> -> -T3a}]
Out[]= $\{\partial_0A, \partial_aA\}$

In[]:= % == {ParamD[TimePar][PPN[A][]], PD[-T3a][PPN[A][]]}
Out[]= True
\end{lstlisting}
Observe that the time derivative is not represented as a partial derivative, but as a parameter derivative with respect to the time parameter \lstinline/TimePar/. Finally, we remark that the automatic split is implemented only for partial derivatives \lstinline/PD[-T4<a>]/; any other derivatives most be converted with \lstinline/ChangeCovD/ or \lstinline/LieDToCovD/, as appropriate.

\subsection{Decomposition into velocity orders}\label{ssec:veldecomp}
As explained in section~\ref{ssec:perturb}, an important part of the PPN formalism is the series expansion of expressions in velocity orders. In order to select a single term \(\order{X}{n} \sim \mathcal{O}(n)\) from an expression \(X\), \xPPN{} provides the function \lstinline/VelocityOrder/. The term \(\order{X}{n}\) is then expressed as \lstinline/VelocityOrder[$X$, $n$]/. Here \(n\) must be a non-negative integer, while \(X\) can be any tensorial expression on \(S_3\), which may in addition depend on \lstinline/TimePar/. It makes use of a number of standard relations for the velocity order in the PPN formalism: orders are distributed over products, and time derivatives are weighted with an additional velocity order. This is illustrated in the following example:
\begin{lstlisting}
In[]:= DefTensor[A[T4<a>], MfSpacetime]
In[]:= VelocityOrder[PPN[A][T3a] PPN[A][-T3a], 2]
Out[]= $\order{A}{0}^a\order{A}{2}_a + \order{A}{1}^a\order{A}{1}_a + \order{A}{2}^a\order{A}{0}_a$

In[]:= VelocityOrder[ParamD[TimePar][PPN[A][-T3a]] + PD[-T3a][PPN[A][-LI[0]]], 3]
Out[]= $\partial_0\order{A}{2}_a + \partial_a\order{A}{3}_0$
\end{lstlisting}
The function \lstinline/VelocityOrder/ supports the boolean option \lstinline/UsePPNRules/. If it is set to \lstinline/True/ (the default case), any rules assigned to \lstinline/PPNTensor/ objects are applied immediately as soon as they are encountered. If it is set to \lstinline/False/, no PPN rules are applied. For example, for the energy-momentum tensor this yields the following results:
\begin{lstlisting}
In[]:= VelocityOrder[PPN[EnergyMomentum][-LI[0], -LI[0]], 2, UsePPNRules -> True]
Out[]= $\rho$

In[]:= % == Density[]
Out[]= True

In[]:= VelocityOrder[PPN[EnergyMomentum][-LI[0], -LI[0]], 2, UsePPNRules -> False]
Out[]= $\order{\Theta}{2}_{00}$

In[]:= % == PPN[EnergyMomentum, 2][-LI[0], -LI[0]]
Out[]= True
\end{lstlisting}
Observe that in the first case the rule \(\order{\Theta}{2}_{00} \to \rho\) is applied, but not in the second case. For large expressions the default setting \lstinline/UsePPNRules -> True/ is significantly faster, since the PPN rules imply the vanishing of many terms in the post-Newtonian expansion, which are thus immediately discarded when the rules are applied.

\subsection{Euler equations}\label{ssec:euler}
An important assumption of the PPN formalism is that the energy-momentum tensor \(\Theta_{\alpha\beta}\) introduced in section~\ref{ssec:enmomdef} satisfies the covariant conservation equation \(\lc{\nabla}_{\alpha}\Theta^{\alpha\beta} = 0\). Expanding this equation into velocity orders and performing a $3+1$ split into space and time components, one obtains the Euler equations, which govern the dynamics of the fluid. These equations are implemented in \xPPN{} in three different functions, which perform the following substitutions on all matching subexpressions of their arguments:
\begin{enumerate}
\item
\lstinline/TimeRhoToEuler[$X$]/ applies the replacement
\begin{equation}
\rho_{,0} \to -(\rho v_a)_{,a}\,.
\end{equation}

\item
\lstinline/TimeVelToEuler[$X$]/ applies the replacement
\begin{equation}
v_{a,0} \to \frac{1}{2}\order{g}{2}_{00,a} - v_bv_{a,b} - \frac{p_{,a}}{\rho}\,.
\end{equation}

\item
\lstinline/TimePiToEuler[$X$]/ applies the replacement
\begin{equation}
\Pi_{,0} \to v_a\left(\frac{p_{,a}}{\rho} - \Pi_{,a} - \frac{1}{2}\order{g}{2}_{00,a} - \frac{1}{2}\order{g}{2}_{bb,a}\right) - \frac{pv_{a,a}}{\rho} - \frac{1}{2}\order{g}{2}_{aa,0}\,.
\end{equation}
\end{enumerate}
The functions can be successively applied in order to transform terms involving higher order time derivatives.

\subsection{Transformation of PPN potentials}\label{ssec:pottrans}
The post-Newtonian potentials defined in section~\ref{ssec:potdef} and their derivatives satisfy a number of relations among each other and with the energy-momentum variables defined in section~\ref{ssec:enmomdef}, some of which follow directly from the definition of the potentials, while others can be derived from the Euler equations discussed in the previous section. \xPPN{} defines a number of utility functions in order to implement these relations and use them to turn PPN potentials and their derivatives into either other PPN potentials or terms constructed from the energy-momentum tensor. The most important of these is \lstinline/PotentialToSource[$X$]/, which performs the following replacements on all subexpressions of \(X\):
\begin{gather}
\triangle\triangle\chi \to 8\pi\rho\,, \quad
\triangle\triangle\mathcal{A} \to 8\pi(\rho v_av_b)_{,ab} - 4\pi\triangle(\rho|v|^2)\,, \quad
\triangle\triangle\mathcal{B} \to 8\pi[\triangle p - (U_{,a}\rho)_{,a}]\,,\nonumber\\
\triangle\Phi_1 \to -4\pi\rho|v|^2\,, \quad
\triangle\Phi_2 \to -4\pi\rho U\,, \quad
\triangle\Phi_3 \to -4\pi\rho\Pi\,, \quad
\triangle\Phi_4 \to -4\pi p\,,\\
\triangle U \to -4\pi\rho\,, \quad
\triangle V_a \to -4\pi\rho v_a\,, \quad
\triangle\Phi_W \to 4\pi\rho U - 4U_{,a}U_{,a} + 2U_{,ab}\chi_{,ab}\,.\nonumber
\end{gather}
The remaining functions act on specific PPN potentials or their derivatives. Their effects are listed in table~\ref{tab:pottrans}.

\begin{table}[htb]
\centering
\begin{tabular}{l|c}
Function & Transformation\\\hline
\lstinline/PotentialChiToU/ & $\triangle\chi \to -2U$\\
\lstinline/PotentialUToChi/ & $U \to -\frac{1}{2}\triangle\chi$\\
\lstinline/PotentialUToUU/ & $U \to U_{aa}$\\
\lstinline/PotentialUUToU/ & $U_{aa} \to U$\\
\lstinline/PotentialUUToChi/ & $U_{ab} \to \chi_{,ab} - \frac{1}{2}\triangle\chi\delta_{ab}$\\
\lstinline/PotentialUToV/ & $U_{,0} \to -V_{a,a}$\\
\lstinline/PotentialUToW/ & $U_{,0} \to W_{a,a}$\\
\lstinline/PotentialVToU/ & $V_{a,a} \to -U_{,0}$\\
\lstinline/PotentialWToU/ & $W_{a,a} \to U_{,0}$\\
\lstinline/PotentialVToW/ & $V_{a,a} \to -W_{a,a}$\\
\lstinline/PotentialWToV/ & $W_{a,a} \to -V_{a,a}$\\
\lstinline/PotentialVToChiW/ & $V_a \to W_a + \chi_{,0a}$\\
\lstinline/PotentialWToChiV/ & $W_a \to V_a - \chi_{,0a}$\\
\lstinline/PotentialChiToPhiAB/ & $\chi_{,00} \to \mathcal{A} + \mathcal{B} - \Phi_1$\\
\lstinline/PotentialUToPhiAB/ & $U_{,00} \to -\frac{1}{2}\triangle(\mathcal{A} + \mathcal{B} - \Phi_1)$
\end{tabular}
\caption{Functions to transform specific PPN potentials.}
\label{tab:pottrans}
\end{table}

\subsection{Sorting of derivatives}\label{ssec:derivsort}
In \xAct, derivatives are, in general, not sorted automatically. This poses two difficulties, which may lead to problems in simplifications:
\begin{enumerate}
\item
Mathematica does not recognize terms which are mathematically equal if they contain derivatives in different order. For example, expressions of the form \(\partial_{\alpha}\partial_{\beta}X\) and \(\partial_{\beta}\partial_{\alpha}X\) are equal, since partial derivatives commute, but since they are represented differently, Mathematica does not recognize this.

\item
In order to apply the transformations listed in section~\ref{ssec:pottrans}, pattern matching is applied to find divergences, time derivatives and Laplace operators acting on tensors. However, due to the way how derivatives are represented in \xTensor, Mathematica recognizes such terms only if they are not interspersed with other derivatives. For example, \(\partial_{\alpha}A^{\alpha}\) is found in \(\partial_{\beta}\partial_{\alpha}A^{\alpha}\), but not in \(\partial_{\alpha}\partial_{\beta}A^{\alpha}\), even though these terms are mathematically equal.
\end{enumerate}

The first of these problems can be solved by defining a canonical order for derivatives and keeping them always sorted in canonical order. Such a canonical ordering is implemented as part of the \xTensor{} function \lstinline/ToCanonical/, so that the following terms are recognized by Mathematica as equal and canceled:

\begin{lstlisting}
In[]:= DefTensor[A[], MfSpacetime]

In[]:= PD[-T4<a>][PD[-T4<b>][A[]]] - PD[-T4<b>][PD[-T4<a>][A[]]]
Out[]= $\partial_{\alpha}\partial_{\beta}A - \partial_{\beta}\partial_{\alpha}A$

In[]:= ToCanonical[%]
Out[]= $0$
\end{lstlisting}

However, this does not solve the second problem, since a different order of derivatives is required, depending on the type of pattern to be matched; for example, matching a time derivative would require the order \(\partial_{\alpha}\partial_0A^{\alpha}\), while matching a divergence would require the order \(\partial_0\partial_{\alpha}A^{\alpha}\). Hence, keeping all indices in a fixed, canonical order would not allow for such patterns to be found. Hence, \xPPN{} implements a different method by defining different functions, which allow sorting derivatives on demand in any of the necessary orders. The following functions are defined:

\begin{enumerate}
\item
\lstinline/SortPDs[$X$]/ sorts all derivatives in canonical order by applying the following rules:
\begin{enumerate}
\item
Spatial derivatives are applied before time derivatives, i.e., moved to the right.
\item
Spatial derivatives are sorted in lexicographic order, i.e., indices which come earlier in lexicographic order, are applied first.
\item
Derivatives with upper indices are applied before derivatives with otherwise identical lower indices.
\end{enumerate}

\item
\lstinline/SortPDsToTime[$X$, $h$]/ sorts derivatives acting on tensors with head \(h\) such that time derivatives are applied first, i.e., moved to the right.

\item
\lstinline/SortPDsToDiv[$X$, $h$]/ sorts derivatives acting on tensors with head \(h\) such that divergences may be matched, i.e., such that derivatives are applied first, whose index is contracted with a corresponding index of the tensor expression.

\item
\lstinline/SortPDsToBox[$X$, $h$]/ sorts derivatives acting on tensors with head \(h\) such that Laplace operators are formed from spatial derivatives, if possible, and those are applied first.
\end{enumerate}

The effect of these functions on an expression with multiple derivatives acting on a tensor is shown by the following code:

\begin{lstlisting}
In[]:= DefTensor[A[T4<a>], MfSpacetime]

In[]:= expr = PD[T3b][ParamD[TimePar][PD[-T3a][PD[-T3c][PD[-T3b][PPN[A][T3c]]]]]]
Out[]= $\partial^b\partial_0\partial_a\partial_c\partial_bA^c$

In[]:= SortPDs[expr]
Out[]= $\partial_0\partial_c\partial_b\partial^b\partial_aA^c$

In[]:= SortPDsToTime[expr, A]
Out[]= $\partial^b\partial_a\partial_c\partial_b\partial_0A^c$

In[]:= SortPDsToDiv[expr, A]
Out[]= $\partial^b\partial_0\partial_a\partial_b\partial_cA^c$

In[]:= SortPDsToBox[expr, A]
Out[]= $\partial_0\partial_a\partial_c\partial_b\partial^bA^c$
\end{lstlisting}

The implementation of these functions is inspired by the field theory package \xTras{} \cite{Nutma:2013zea}, which defines a similar set of functions, which take into account also non-commuting derivatives by adding suitable correction terms. Also note that the replacement functions listed in section~\ref{ssec:pottrans} make use of these sorting functions automatically in order to establish the necessary order of indices before pattern matching is performed.

\section{Example: scalar-tensor gravity}\label{sec:example}
In order to demonstrate the use of the \xPPN{} package, we provide a practical example in form of a complete, commented session, which calculates the PPN parameters of a scalar-tensor class of gravity theories, thereby showing the use of some of the functions detailed in the previous sections. The precise steps which are needed to determine the PPN parameters highly depend on the specific theory under consideration, and must be chosen accordingly. We briefly present the action and field equations of this class in section~\ref{ssec:exaction}. A few preliminary commands, for loading the package and setup, are given in section~\ref{ssec:loading}. In section~\ref{ssec:defi}, the necessary tensors and constants for the calculation are defined. These are used in the definition of the field equations in section~\ref{ssec:fieldeq}. Their post-Newtonian expansion is derived in section~\ref{ssec:ppnexp}. In section~\ref{ssec:solution}, the perturbative solution of the resulting equations is obtained. Finally, in section~\ref{ssec:ppnmetric}, the PPN metric and parameters are calculated.

\subsection{Action and field equations}\label{ssec:exaction}
In the following we discuss a class of scalar-tensor theories of gravity, whose action is given by~\cite{Nordtvedt:1970uv}
\begin{equation}\label{eq:exaction}
S = \frac{1}{2\kappa^2}\int_{M_4}\dd^4x\sqrt{-g}\left(\psi R - \frac{\omega(\psi)}{\psi}\partial_{\rho}\psi\partial^{\rho}\psi\right) + S_m[g_{\mu\nu}, \chi]
\end{equation}
in Brans-Dicke like parametrization in the Jordan conformal frame. Here \(S_m\) denotes the matter part of the action, where we collectively denoted by \(\chi\) the set of matter fields. The gravitational part contains a free function \(\omega\) of the scalar field \(\psi\). Each theory of this class is defined by a particular choice of this free function \(\omega\). By variation of this action with respect to the metric and the scalar field as well as subtraction of a suitable multiple of the trace of the metric field equation one obtains the field equations
\begin{subequations}\label{eq:stgfeq}
\begin{align}
\psi R_{\mu\nu} - \lc{\nabla}_{\mu}\partial_{\nu}\psi - \frac{\omega}{\psi}\partial_{\mu}\psi\partial_{\nu}\psi + \frac{g_{\mu\nu}}{4\omega + 6}\frac{d\omega}{d\psi}\partial_{\rho}\psi\partial^{\rho}\psi &= \kappa^2\left(\Theta_{\mu\nu} - \frac{\omega + 1}{2\omega + 3}g_{\mu\nu}\Theta\right)\,,\label{eq:stgmeteq}\\
(2\omega + 3)\lc{\square}\psi + \frac{d\omega}{d\psi}\partial_{\rho}\psi\partial^{\rho}\psi &= \kappa^2\Theta\,,\label{eq:stgscaleq}
\end{align}
\end{subequations}
where \(\lc{\square} = g^{\mu\nu}\lc{\nabla}_{\mu}\lc{\nabla}_{\nu}\) is the d'Alembert operator on the physical spacetime \(M_4\).

\subsection{Package loading and preliminaries}\label{ssec:loading}
In order to load and use \xPPN, as the first prerequisite a working installation of \xAct{} \cite{xact} is needed. The files provided by \xAct{} then reside in a directory named \texttt{xAct} in the Mathematica package search path. The \xPPN{} package comes as a directory named \texttt{xPPN}, which must be placed inside the \texttt{xAct} directory. If both packages are installed correctly, \xPPN{} can be loaded with the command
\begin{lstlisting}
In[]:= << xAct`xPPN`
\end{lstlisting}
This should load \xPPN{} and its dependencies from the \xAct{} package suite. Note that loading may take some time, since \xPPN{} calculates the perturbative expansion of the pre-defined tensor fields upon package loading. To suppress \$ symbols in the index notation, it is useful to set the following \xAct{} printing option:
\begin{lstlisting}
In[]:= §PrePrint = ScreenDollarIndices;
\end{lstlisting}
Finally, we define two utility functions which help to create rules from equations; this functionality, which is simply a shorthand notation for the versatile function \lstinline/MakeRule/ from the \xTensor{} package, are not specific to \xPPN, and are defined here only to shorten the notation in the later course of this example:
\begin{lstlisting}
In[]:= mkrg[eq_Equal] :=  MakeRule[Evaluate[List @@ eq],
	MetricOn -> All, ContractMetrics -> True]

In[]:= mkr0[eq_Equal] :=  MakeRule[Evaluate[List @@ eq],
	MetricOn -> None, ContractMetrics -> False]
\end{lstlisting}

\subsection{Object definitions}\label{ssec:defi}
We continue by defining the \xAct{} objects which we will be using for the calculation and their notation. We start with the scalar field \(\psi\), which is a tensor without any indices.
\begin{lstlisting}
In[]:= DefTensor[psi[], MfSpacetime, PrintAs -> "$\psi$"]
\end{lstlisting}
We then continue with the cosmological background value \(\Psi\) of the scalar field. This is a constant.
\begin{lstlisting}
In[]:= DefConstantSymbol[psi0, PrintAs -> "$\Psi$"]
\end{lstlisting}
Another constant which we will need is the gravitational constant \(\kappa\).
\begin{lstlisting}
In[]:= DefConstantSymbol[kappa, PrintAs -> "$\kappa$"]
\end{lstlisting}
The action~\eqref{eq:exaction} also depends on a free function \(\omega\). This is defined as a scalar function.
\begin{lstlisting}
In[]:= DefScalarFunction[omega, PrintAs -> "$\omega$"]
\end{lstlisting}
We then come to the field equations~\eqref{eq:stgfeq}, which we write in the form \(\mathcal{E}_{\alpha\beta} = 0\) and \(\mathcal{E} = 0\), by moving the energy-momentum tensor to the left hand side. The two tensors representing these field equations are then defined as follows.
\begin{lstlisting}
In[]:= DefTensor[MetEq[-T4<a>, -T4<b>], MfSpacetime, Symmetric[{1, 2}], PrintAs -> "$\mathcal{E}$"]
In[]:= DefTensor[ScalEq[], MfSpacetime, PrintAs -> "$\mathcal{E}$"]
\end{lstlisting}
Finally, in this example we will solve the field equations by making an ansatz for the solution in terms of PPN potentials and unknown, constant coefficients, which we will then determine. For this purpose, we define a function which creates such constant coefficients on demand as follows.
\begin{lstlisting}
In[]:= aa[i_] := Module[{sym = Symbol["a" <> ToString[i]]},
	If[!ConstantSymbolQ[sym],
		DefConstantSymbol[sym, PrintAs -> StringJoin["\!\(a\_", ToString[i], "\)"]]
	];
	Return[sym]]
\end{lstlisting}

\subsection{Field equations}\label{ssec:fieldeq}
In the next step, we enter the field equations~\eqref{eq:stgfeq}. As mentioned before, we must pay attention to explicitly use the inverse metric \(g^{\alpha\beta}\) for terms such as the kinetic term \(\partial_{\rho}\psi\partial^{\rho}\psi\) of the scalar field or the trace \(\Theta^{\rho}{}_{\rho}\) of the energy-momentum tensor. Taking these into account, we can enter the metric field equation~\eqref{eq:stgmeteq} as follows.
\begin{lstlisting}
In[]:= psi[] * RicciCD[-T4<a>, -T4<b>] - CD[-T4<a>][CD[-T4<b>][psi[]]] -
	PD[-T4<a>][psi[]] * PD[-T4<b>][psi[]] * omega[psi[]] / psi[] +
	InvMet[T4<g>, T4<d>] * PD[-T4<g>][psi[]] * PD[-T4<d>][psi[]] * Met[-T4<a>, -T4<b>] *
	omega'[psi[]] / (4 omega[psi[]] + 6) -
	(EnergyMomentum[-T4<a>, -T4<b>] - InvMet[T4<g>, T4<d>] * EnergyMomentum[-T4<g>, -T4<d>] *
	Met[-T4<a>, -T4<b>] * (omega[psi[]] + 1) / (2 omega[psi[]] + 3)) * kappa^2;

In[]:= meteqdef = MetEq[-T4<a>, -T4<b>] == %;
In[]:= meteqru = mkr0[meteqdef];
\end{lstlisting}
Similarly, we continue with the scalar field equation~\eqref{eq:stgscaleq}:
\begin{lstlisting}
In[]:= (2 omega[psi[]] + 3) * InvMet[T4<a>, T4<b>] * CD[-T4<a>][CD[-T4<b>][psi[]]] +
	omega'[psi[]] * InvMet[T4<a>, T4<b>] * PD[-T4<a>][psi[]] * PD[-T4<b>][psi[]] -
	kappa^2 * InvMet[T4<a>, T4<b>] * EnergyMomentum[-T4<a>, -T4<b>];

In[]:= scaleqdef = ScalEq[] == %;
In[]:= scaleqru = mkr0[scaleqdef];
\end{lstlisting}

\subsection{Post-Newtonian expansion}\label{ssec:ppnexp}
The most crucial and resource intensive part of the calculation is the derivation of the post-Newtonian expansion of the field equations. For this purpose, we must first define the expansion of the scalar field. Here we impose the relations
\begin{equation}
\order{\psi}{0} = \Psi\,, \quad
\order{\psi}{1} = 0\,, \quad
\order{\psi}{3} = 0\,.
\end{equation}
In \xPPN, they are implemented as follows:
\begin{lstlisting}
In[]:= OrderSet[PPN[psi, 0][], psi0];
In[]:= OrderSet[PPN[psi, 1][], 0];
In[]:= OrderSet[PPN[psi, 3][], 0];
\end{lstlisting}
With these definitions in place, we can continue with the calculation. We will do so in two steps. First, we perform a $3+1$ decomposition of the field equations. For this purpose, we convert the Levi-Civita covariant derivatives to partial derivatives and Christoffel symbols. Also we perform a number of simplifications. Again we start with the metric field equations.
\begin{lstlisting}
In[]:= {#, # /. meteqru} &[MetEq[-T4<a>, -T4<b>]];
In[]:= ChangeCovD[%, CD, PD];
In[]:= Expand[%];
In[]:= SpaceTimeSplits[#, {-T4<a> -> -T3a, -T4<b> -> -T3b}] & /@ %;
In[]:= Expand[%];
In[]:= Map[ToCanonical, %, {3}];
In[]:= Map[SortPDs, %, {3}];
In[]:= meteq31list = %;
In[]:= meteq31def = Union[Flatten[MapThread[Equal, %, 2]]];
In[]:= meteq31ru = Flatten[mkrg /@ %];
\end{lstlisting}
We proceed similarly with the scalar field equations:
\begin{lstlisting}
In[]:= {#, # /. scaleqru} &[ScalEq[]];
In[]:= ChangeCovD[%, CD, PD];
In[]:= Expand[%];
In[]:= SpaceTimeSplit[#, {}] & /@ %;
In[]:= Expand[%];
In[]:= ToCanonical /@ %;
In[]:= SortPDs /@ %;
In[]:= scaleq31list = %;
In[]:= scaleq31def = Equal @@ %;
In[]:= scaleq31ru = Flatten[mkrg[%]];
\end{lstlisting}
In the second step, we further decompose the equations obtained in the previous step into velocity orders \(\mathcal{O}(0), \ldots, \mathcal{O}(4)\). Also here we perform a few tensor simplifications alongside the calculation, starting again with the metric equation.
\begin{lstlisting}
In[]:= Outer[VelocityOrder, meteq31list, Range[0, 4]];
In[]:= Map[NoScalar, %, {4}];
In[]:= Expand[%];
In[]:= Map[ContractMetric[#, OverDerivatives -> True,
	AllowUpperDerivatives -> True] &, %, {4}];
In[]:= Map[ToCanonical, %, {4}];
In[]:= Map[SortPDs, %, {4}];
In[]:= meteqvlist = Simplify[%];
In[]:= meteqvdef = Union[Flatten[MapThread[Equal, %, 3]]]
In[]:= meteqvru = Flatten[mkrg /@ %];
\end{lstlisting}
Finally, we apply the same step also to the scalar field equation.
\begin{lstlisting}
In[]:= Outer[VelocityOrder, scaleq31list, Range[0, 4]];
In[]:= Map[NoScalar, %, {2}];
In[]:= Expand[%];
In[]:= Map[ContractMetric[#, OverDerivatives -> True,
	AllowUpperDerivatives -> True] &, %, {2}];
In[]:= Map[ToCanonical, %, {2}];
In[]:= Map[SortPDs, %, {2}];
In[]:= scaleqvlist = Simplify[%];
In[]:= scaleqvdef = Flatten[MapThread[Equal, %, 1]]
In[]:= scaleqvru = Flatten[mkrg /@ %];
\end{lstlisting}

\subsection{Perturbative solution}\label{ssec:solution}
We can now use the post-Newtonian expansion of the field equations and solve them order by order. For this purpose we make use of prior knowledge that the field equations of the theory at hand can be solved by a particular ansatz for the metric and scalar field, so that at every step of the calculation we are left with solving for a number of constant coefficients; this ansatz and solution method must be adapted if the package is applied to other theories. We show the zeroth velocity order (the background vacuum solution) in section~\ref{sssec:vel0}, the second velocity order in section~\ref{sssec:vel2}, the third velocity order in section~\ref{sssec:vel3} and the fourth velocity order in section~\ref{sssec:vel4}.

\subsubsection{Zeroth velocity order}\label{sssec:vel0}
First, we verify that the background vacuum equations are solved identically by the background geometry. This can be checked as follows.
\begin{lstlisting}
In[]:= PPN[MetEq, 0][-LI[0], -LI[0]] /. meteqvru
Out[]= $0$

In[]:= PPN[MetEq, 0][-T3a, -T3b] /. meteqvru
Out[]= $0$

In[]:= PPN[ScalEq, 0][] /. scaleqvru
Out[]= $0$
\end{lstlisting}

\subsubsection{Second velocity order}\label{sssec:vel2}
We then continue with the second velocity order. First, we gather the equations \(\order{\mathcal{E}}{2}_{00}, \order{\mathcal{E}}{2}_{ab}, \order{\mathcal{E}}{2}\) to be solved:
\begin{lstlisting}
In[]:= eqs2 = FullSimplify[{PPN[MetEq, 2][-LI[0], -LI[0]],
	PPN[MetEq, 2][-T3a, -T3b], PPN[ScalEq, 2][]} /. meteqvru /. scaleqvru];
\end{lstlisting}
These equations take the form
\begin{gather}
\order{\mathcal{E}}{2} = \kappa^2\rho + (2\omega(\Psi) + 3)\triangle\order{\psi}{2}\,, \quad
\order{\mathcal{E}}{2}_{00} = -\kappa^2\rho\frac{\omega(\Psi) + 2}{2\omega(\Psi) + 3} - \frac{\Psi}{2}\triangle\order{g}{2}_{00}\,,\nonumber\\
\order{\mathcal{E}}{2}_{ab} = -\kappa^2\rho\frac{\omega(\Psi) + 1}{2\omega(\Psi) + 3}\delta_{ab} + \frac{\Psi}{2}\left(\order{g}{2}_{00,ab} - \order{g}{2}_{cc,ab} + 2\order{g}{2}_{c(a,b)c} - \triangle\order{g}{2}_{ab}\right) - \order{\psi}{2}_{,ab}\,.
\end{gather}
The easiest way to solve these equations is using an ansatz for the metric and scalar field perturbations in terms of the post-Newtonian potentials, with arbitrary, constant coefficients. This ansatz can be defined as follows:
\begin{lstlisting}
In[]:= ans2def = {PPN[Met, 2][-LI[0], -LI[0]] == aa[1] * PotentialU[],
	PPN[Met, 2][-T3a, -T3b] == aa[2] * PotentialU[] * BkgMetricS3[-T3a, -T3b] +
		aa[3] * PotentialUU[-T3a, -T3b],
	PPN[psi, 2][] == aa[4] * PotentialU[]}
Out[]= $\big\{\order{g}{2}_{00} = a_1U\,, \quad \order{g}{2}_{ab} = a_2U\delta_{ab} + a_3U_{ab}\,, \quad \order{\psi}{2} = a_4U\big\}$

In[]:= ans2ru = Flatten[mkrg /@ ans2def];
\end{lstlisting}
To obtain the equations to be solved, one inserts this ansatz into the field equations. Since the field equations contain derivatives of the metric and scalar field perturbations, this will yield derivatives of the post-Newtonian potentials \(U\) and \(U_{ab}\). These must be matched with the terms arising from the energy-momentum tensor \(\Theta_{\alpha\beta}\). For this purpose, one applies the functions listed in section~\ref{ssec:pottrans}. For the potentials at hand it is most convenient to first convert them to derivatives of the superpotential \(\chi\), before transforming them to terms involving the matter density \(\rho\).
\begin{lstlisting}
In[]:= eqs2 /. ans2ru;
In[]:= PotentialUToChi /@ %;
In[]:= PotentialUUToChi /@ %;
In[]:= Expand[%];
In[]:= ToCanonical /@ %;
In[]:= ContractMetric[#, OverDerivatives -> True, AllowUpperDerivatives -> True] & /@ %;
In[]:= PotentialToSource /@ %;
In[]:= Expand[%];
In[]:= ToCanonical /@ %;
In[]:= SortPDs /@ %;
In[]:= eqsa2 = FullSimplify[%];
\end{lstlisting}
Inspecting the obtained equations shows that they are given by
\begin{gather}
\order{\mathcal{E}}{2} = \kappa^2\rho - 4\pi a_4(2\omega(\Psi) + 3)\rho\,, \quad
\order{\mathcal{E}}{2}_{00} = 2\pi\Psi a_1\rho - \kappa^2\rho\frac{\omega(\Psi) + 2}{2\omega(\Psi) + 3}\,,\nonumber\\
\order{\mathcal{E}}{2}_{ab} = \left(\frac{a_4}{2} + \frac{a_2 + a_3 - a_1}{4}\Psi\right)\triangle\chi_{,ab} + \left[2\pi\Psi(a_2 + a_3) - \kappa^2\frac{\omega(\Psi) + 1}{2\omega(\Psi) + 3}\right]\rho\delta_{ab}\,.
\end{gather}
These equations are solved if and only if the coefficients of each of the terms \(\rho\), \(\rho\delta_{ab}\) and \(\triangle\chi_{,ab}\) vanishes individually. This yields four equations for the four coefficients \(a_1, \ldots, a_4\). However, one finds that these are not linearly independent, due to the gauge freedom arising from the diffeomorphism invariance of the theory. Hence, they must be supplemented with an additional, gauge fixing equation. The common choice in the PPN formalism is to eliminate the term \(U_{ab}\) from the component \(\order{g}{2}_{ab}\), thus setting \(a_3 = 0\). We can thus determine the component equations:
\begin{lstlisting}
In[]:= eqsc2 = FullSimplify[{
	Coefficient[eqsa2[[1]], Density[]], Coefficient[eqsa2[[3]], Density[]],
	Coefficient[eqsa2[[2]], Density[] * BkgMetricS3[-T3a, -T3b]], aa[3]}];
\end{lstlisting}
We can then solve the equations:
\begin{lstlisting}
In[]:= sola2 = FullSimplify[First[Solve[# == 0 & /@ eqsc2, aa /@ Range[1, 4]]]];
\end{lstlisting}
This yields the solution
\begin{equation}
a_1 = \kappa^2\frac{\omega(\Psi) + 2}{2\pi\Psi(2\omega(\Psi) + 3)}\,, \quad
a_2 = \kappa^2\frac{\omega(\Psi) + 1}{2\pi\Psi(2\omega(\Psi) + 3)}\,, \quad
a_3 = 0\,, \quad
a_4 = \frac{\kappa^2}{4\pi(2\omega(\Psi) + 3)}\,.
\end{equation}
We may check that this indeed solves the equations:
\begin{lstlisting}
In[]:= Simplify[eqsa2 /. sola2]
Out[]= {0, 0, 0}
\end{lstlisting}
Together with the ansatz we defined, this solution now yields the solution for the perturbations \(\order{g}{2}_{00}, \order{g}{2}_{ab}, \order{\psi}{2}\):
\begin{lstlisting}
In[]:= sol2def = ans2def /. sola2;
In[]:= sol2ru = Flatten[mkrg /@ sol2def];
\end{lstlisting}
Finally, we also check that this result solves the second-order field equations:
\begin{lstlisting}
In[]:= eqs2 /. sol2ru;
In[]:= Expand[%];
In[]:= PotentialToSource /@ %;
In[]:= ToCanonical /@ %;
In[]:= SortPDs /@ %;
In[]:= Simplify[%]
Out[]= {0, 0, 0}
\end{lstlisting}

\subsubsection{Third velocity order}\label{sssec:vel3}
We then come to the third velocity order. Also here we proceed in full analogy to the second velocity order shown above. First, we isolate the field equation \(\order{\mathcal{E}}{3}_{0a}\) which we will solve:
\begin{lstlisting}
In[]:= eqs3 = FullSimplify[PPN[MetEq, 3][-LI[0], -T3a] /. meteqvru];
\end{lstlisting}
This equation takes the form
\begin{equation}
\order{\mathcal{E}}{3}_{0a} = \kappa^2\rho v_a - \order{\psi}{2}_{,0a} + \frac{\Psi}{2}\left(\order{g}{3}_{0b,ab} - \triangle\order{g}{3}_{0a} + \order{g}{2}_{ab,0b} - \order{g}{2}_{bb,0a}\right)\,.
\end{equation}
We then choose an ansatz for the third-order metric perturbation \(\order{g}{3}_{0a}\). This now involves the post-Newtonian vector potentials \(V_a\) and \(W_a\) with constant coefficients:
\begin{lstlisting}
In[]:= ans3def = PPN[Met, 3][-LI[0], -T3a] ==
	aa[5] * PotentialV[-T3a] + aa[6] * PotentialW[-T3a]
Out[]= $\order{g}{3}_{0a} = a_5V_a + a_6W_a$

In[]:= ans3ru = mkrg[ans3def];
\end{lstlisting}
To obtain the equation to be solved, we insert this ansatz into the field equation, together with the second-order solution obtained in the previous step. Also here we obtain derivatives acting on the post-Newtonian potentials, which we can simplify by using their interrelations, and finally convert them to terms matching the energy-momentum source. Here it is most useful to first eliminate the potential \(W_a\). The resulting terms can then be converted to source terms and terms involving derivatives acting on \(U\) only:
\begin{lstlisting}
In[]:= eqs3 /. ans3ru /. sol2ru;
In[]:= PotentialWToChiV[%];
In[]:= Expand[%];
In[]:= ContractMetric[%, OverDerivatives -> True, AllowUpperDerivatives -> True];
In[]:= PotentialChiToU[%];
In[]:= PotentialVToU[%];
In[]:= PotentialToSource[%];
In[]:= ToCanonical[%];
In[]:= SortPDs[%];
In[]:= eqsa3 = FullSimplify[%];
\end{lstlisting}
Inspecting this equation shows the following form:
\begin{equation}
\order{\mathcal{E}}{3}_{0a} = [\kappa^2 + 2\pi\Psi(a_5 + a_6)]\left(\rho v_a - \frac{U_{,0a}}{4\pi}\right)\,.
\end{equation}
This equation only determines the sum \(a_5 + a_6\) of the coefficients we introduced. Here we leave their difference as an undetermined parameter, which we will solve for when we come to the fourth velocity order, which will fix the post-Newtonian gauge. We thus determine the solution:
\begin{lstlisting}
In[]:= sola3 = FullSimplify[
	First[Solve[{eqsa3 == 0, aa[6] - aa[5] == aa[0]}, {aa[5], aa[6]}]]];
\end{lstlisting}
The solution is given by
\begin{equation}
a_5 = -\frac{a_0}{2} - \frac{\kappa^2}{4\pi\Psi}\,, \quad
a_6 = \frac{a_0}{2} - \frac{\kappa^2}{4\pi\Psi}\,.
\end{equation}
We check that this solves the equations:
\begin{lstlisting}
In[]:= Simplify[eqsa3 /. sola3]
Out[]= $0$
\end{lstlisting}
Together with the ansatz for the third-order metric component \(\order{g}{3}_{0a}\) we obtain the solution:
\begin{lstlisting}
In[]:= sol3def = ans3def /. sola3;
In[]:= sol3ru = mkrg[sol3def];
\end{lstlisting}
Finally, we also check that this solves the third-order field equation we started with:
\begin{lstlisting}
In[]:= eqs3 /. sol2ru /. sol3ru;
In[]:= PotentialWToChiV[%];
In[]:= Expand[%];
In[]:= ContractMetric[%, OverDerivatives -> True, AllowUpperDerivatives -> True];
In[]:= PotentialChiToU[%];
In[]:= PotentialVToU[%];
In[]:= PotentialToSource[%];
In[]:= ToCanonical[%];
In[]:= SortPDs[%];
In[]:= Simplify[%]
Out[]= $0$
\end{lstlisting}

\subsubsection{Fourth velocity order}\label{sssec:vel4}
Finally, we come to the fourth velocity order, which is the most involved. For the theory under consideration, it is sufficient to solve the field equation \(\order{\mathcal{E}}{4}_{00}\), as it determines the necessary component \(\order{g}{4}_{00}\). We hence isolate this equation:
\begin{lstlisting}
In[]:= eqs4 = PPN[MetEq, 4][-LI[0], -LI[0]] /. meteqvru;
\end{lstlisting}
We find that it takes the following form:
\begin{multline}
\order{\mathcal{E}}{4}_{00} = -\frac{\Psi}{4}\left(2\triangle\order{g}{4}_{00} - 4\order{g}{3}_{0a,0a} + 2\order{g}{2}_{aa,00} + \order{g}{2}_{00,a}\order{g}{2}_{00,a} + \order{g}{2}_{00,a}\order{g}{2}_{bb,a} - 2\order{g}{2}_{00,a}\order{g}{2}_{ab,b} - \order{g}{2}_{00,ab}\order{g}{2}_{ab}\right)\\
- \frac{\omega'(\Psi)}{4\omega(\Psi) + 6}\order{\psi}{2}_{,a}\order{\psi}{2}_{,a} + \frac{\kappa^2\omega'(\Psi)}{(2\omega(\Psi) + 3)^2}\rho\order{\psi}{2} + \kappa^2\frac{\omega(\Psi) + 2}{2\omega(\Psi) + 3}\rho\order{g}{2}_{00} - \frac{1}{2}\order{\psi}{2}_{,a}\order{g}{2}_{00,a} - \frac{1}{2}\order{\psi}{2}\triangle\order{g}{2}_{00} - \order{\psi}{2}_{,00}\\
- \kappa^2\rho|v|^2 - \kappa^2\frac{\omega(\Psi) + 2}{2\omega(\Psi) + 3}\rho\Pi - 3\kappa^2\frac{\omega(\Psi) + 3}{2\omega(\Psi) + 3}p\,.
\end{multline}
As we shall see below, this equation can be solved by the following ansatz:
\begin{lstlisting}
In[]:= ans4def =  PPN[Met, 4][-LI[0], -LI[0]] == aa[11] * PotentialU[]^2 +
	aa[7] * PotentialPhi1[] + aa[8] * PotentialPhi2[] +
	aa[9] * PotentialPhi3[] + aa[10] * PotentialPhi4[]
Out[]= $\order{g}{4}_{00} = a_7\Phi_1 + a_8\Phi_2 + a_9\Phi_3 + a_{10}\Phi_4 + a_{11}U^2$

In[]:= ans4ru = mkrg[ans4def];
\end{lstlisting}
We then insert this ansatz into the fourth-order field equation, alongside the previously determined solutions at lower velocity order. In order to convert the post-Newtonian potentials to a unified form, from which we can read off the independent equations for the constant coefficients in the ansatz, it is sufficient to replace the divergence terms \(V_{a,a}\) and \(W_{a,a}\) by time derivatives of \(U\), before replacing all potentials by source terms. This is done as follows:
\begin{lstlisting}
In[]:= eqs4 /. ans4ru /. sol2ru /. sol3ru;
In[]:= Expand[%];
In[]:= ContractMetric[%, OverDerivatives -> True, AllowUpperDerivatives -> True];
In[]:= PotentialVToU[%];
In[]:= PotentialWToU[%];
In[]:= PotentialToSource[%];
In[]:= ToCanonical[%];
In[]:= SortPDs[%];
In[]:= Expand[%];
In[]:= eqsa4 = Simplify[ScreenDollarIndices[%]];
\end{lstlisting}
The resulting equation \(\order{\mathcal{E}}{4}_{00}\), which we do not display here for brevity, involves the six independent terms \(p\), \(\rho\Pi\), \(\rho U\), \(U_{,00}\), \(\rho|v|^2\), \(U_{,a}U_{,a}\), whose constant coefficients must vanish. We thus isolate these constant coefficients:
\begin{lstlisting}
In[]:= eq1 = Simplify[Coefficient[eqsa4, Pressure[]]];
In[]:= eq2 = Simplify[Coefficient[eqsa4, Density[] * InternalEnergy[]]];
In[]:= eq3 = Simplify[Coefficient[eqsa4, Density[] * PotentialU[]]];
In[]:= eq4 = Simplify[Coefficient[eqsa4, ParamD[TimePar, TimePar][PotentialU[]]]];
In[]:= eq5 = Simplify[Coefficient[eqsa4, Density[] * Velocity[-T3a] * Velocity[T3a]]];
In[]:= eq6 = Simplify[Coefficient[eqsa4, PD[-T3a][PotentialU[]] * PD[T3a][PotentialU[]]]];
\end{lstlisting}
For consistency, it is useful to check the completeness of this decomposition:
\begin{lstlisting}
In[]:= Simplify[Pressure[] * eq1 + Density[] * InternalEnergy[] * eq2 +
	Density[] * PotentialU[] * eq3 + ParamD[TimePar, TimePar][PotentialU[]] * eq4 +
	Density[] * Velocity[-T3a] * Velocity[T3a] * eq5 +
	PD[-T3a][PotentialU[]] * PD[T3a][PotentialU[]] * eq6 - eqsa4]
Out[]= $0$
\end{lstlisting}
We have thus obtained six equations for the six remaining unknowns \(a_0, a_6, \ldots, a_{11}\). It turns out that these equations are linearly independent, so that we can solve for all missing constants:
\begin{lstlisting}
In[]:= sola4 = Simplify[First[Solve[# == 0 & /@ {eq1, eq2, eq3, eq4, eq5, eq6},
	aa /@ Prepend[Range[7, 11], 0]]]];
\end{lstlisting}
This yields the solution
\begin{gather}
a_0 = \frac{\kappa^2}{4\pi\Psi}\frac{3\omega(\Psi) + 4}{2\omega(\Psi) + 3}\,, \quad
a_9 = \frac{\kappa^2}{2\pi\Psi}\frac{\omega(\Psi) + 2}{2\omega(\Psi) + 3}\,, \quad
a_8 = \frac{\kappa^4}{16\pi^2\Psi^2}\frac{12 + 38\omega(\Psi) + 32\omega^2(\Psi) + 8\omega^3(\Psi) - \Psi\omega'(\Psi)}{(2\omega(\Psi) + 3)^3}\,,\nonumber\\
a_7 = \frac{\kappa^2}{2\pi\Psi}\,, \quad
a_{10} = \frac{3\kappa^2}{2\pi\Psi}\frac{\omega(\Psi) + 1}{2\omega(\Psi) + 3}\,, \quad
a_{11} = -\frac{\kappa^4}{32\pi^2\Psi^2}\frac{48 + 80\omega(\Psi) + 44\omega^2(\Psi) + 8\omega^3(\Psi) + \Psi\omega'(\Psi)}{(2\omega(\Psi) + 3)^3}\,.
\end{gather}
We check that it is indeed a solution:
\begin{lstlisting}
In[]:= Simplify[eqsa4 /. sola4]
Out[]= $0$
\end{lstlisting}
Since we have now determined \(a_0\), we can complete the partial third-order solution we obtained earlier:
\begin{lstlisting}
In[]:= sol3def = ans3def /. Simplify[sola3 /. sola4];
In[]:= sol3ru = mkrg[sol3def];
\end{lstlisting}
Using the remaining constants \(a_6, \ldots, a_{11}\), we obtain the solution for \(\order{g}{4}_{00}\):
\begin{lstlisting}
In[]:= sol4def = ans4def /. sola4;
In[]:= sol4ru = mkrg[sol4def];
\end{lstlisting}
Again, we check for consistency that this solves the fourth-order field equation:
\begin{lstlisting}
In[]:= eqs4 /. sol2ru /. sol3ru /. sol4ru;
In[]:= Expand[%];
In[]:= ContractMetric[%, OverDerivatives -> True, AllowUpperDerivatives -> True];
In[]:= PotentialVToU[%];
In[]:= PotentialWToU[%];
In[]:= PotentialToSource[%];
In[]:= ToCanonical[%];
In[]:= SortPDs[%];
In[]:= Expand[%];
In[]:= Simplify[%]
Out[]= $0$
\end{lstlisting}

\subsection{PPN metric and parameters}\label{ssec:ppnmetric}
Using the solution we determined so far, we can now calculate the post-Newtonian metric and parameters. For this purpose, we first isolate the metric components to be determined.
\begin{lstlisting}
In[]:= metcomp = {PPN[Met, 2][-LI[0], -LI[0]], PPN[Met, 2][-T3a, -T3b],
	PPN[Met, 3][-LI[0], -T3a], PPN[Met, 4][-LI[0], -LI[0]]}
Out[]= $\big\{\order{g}{2}_{00}, \order{g}{2}_{ab}, \order{g}{3}_{0a}, \order{g}{4}_{00}\big\}$
\end{lstlisting}
We then insert the solution we have found.
\begin{lstlisting}
In[]:= metcomp /. sol2ru /. sol3ru /. sol4ru;
In[]:= ToCanonical[%];
In[]:= Expand[%];
In[]:= ppnmet = Simplify[%];
\end{lstlisting}
We will compare this to the standard PPN metric.
\begin{lstlisting}
In[]:= stamet = Simplify[MetricToStandard /@ metcomp];
\end{lstlisting}
This can be done in two steps. First, we choose the normalization constant \(\kappa\), which corresponds to the effective Newtonian constant, such that \(\order{g}{2}_{00} = 2U\). We define the following equation:
\begin{lstlisting}
In[]:= kappaeq = First[ppnmet] == First[stamet];
\end{lstlisting}
This takes the form
\begin{equation}
2U = \frac{\kappa^2}{2\pi\Psi}\frac{\omega(\Psi) + 2}{2\omega(\Psi) + 3}U\,.
\end{equation}
To solve this equation, we take its positive root:
\begin{lstlisting}
In[]:= First[Sqrt[FullSimplify[k2 /. Solve[kappaeq /. kappa -> Sqrt[k2], k2]]]];
In[]:= kappadef =  kappa == %;
In[]:= kapparu = mkrg[kappadef];
\end{lstlisting}
This yields the solution
\begin{equation}
\kappa = \sqrt{4\pi\Psi\frac{2\omega(\Psi) + 3}{\omega(\Psi) + 2}}\,.
\end{equation}
With this solution, we can now determine the PPN parameters. These are the equations to be solved.
\begin{lstlisting}
In[]:= pareqs = Simplify[ToCanonical[stamet - ppnmet /. kapparu]];
\end{lstlisting}
To determine the PPN parameters, we isolate the constant coefficients in front of the post-Newtonian potentials \(U\delta_{ab}, V_a, W_a, \mathcal{A}, U^2, \Phi_W, \Phi_1, \Phi_2, \Phi_3, \Phi_4\).
\begin{lstlisting}
In[]:= pots = {PotentialU[] BkgMetricS3[-T3a, -T3b], PotentialV[-T3a], PotentialW[-T3a],
	PotentialA[], PotentialU[]^2, PotentialPhiW[],
	PotentialPhi1[], PotentialPhi2[], PotentialPhi3[], PotentialPhi4[]};
In[]:= eqs = DeleteCases[Flatten[Simplify[Outer[Coefficient, pareqs, pots]]], 0];
\end{lstlisting}
Finally, we can solve for the full set of PPN parameters.
\begin{lstlisting}
In[]:= pars = {ParameterBeta, ParameterGamma, ParameterXi,
	ParameterAlpha1, ParameterAlpha2, ParameterAlpha3,
	ParameterZeta1, ParameterZeta2, ParameterZeta3, ParameterZeta4};
In[]:= parsol = FullSimplify[Solve[# == 0 & /@ eqs, pars][[1]]];
\end{lstlisting}
This finally yields the solution
\begin{equation}\label{eq:exppnpar}
\gamma = \frac{\omega(\Psi) + 1}{\omega(\Psi) + 2}\,, \quad
\beta = 1 + \frac{\Psi\omega'(\Psi)}{4(2\omega(\Psi) + 3)(\omega(\Psi) + 2)^2}\,, \quad
\alpha_1 = \alpha_2 = \alpha_3 = \zeta_1 = \zeta_2 = \zeta_3 = \zeta_4 = \xi = 0\,.
\end{equation}
This is of course the well-known post-Newtonian limit of scalar-tensor gravity with a massless scalar field~\cite{Nordtvedt:1970uv}.

\section{Summary and outlook}\label{sec:conclusion}
We have presented the Mathematica package \xPPN, which is based on the tensor algebra suite \xAct{} and which implements the PPN formalism in its formulation presented in~\cite{Will:1993ns}, with extensions to adapt it to the geometries employed in the three formulations of general relativity~\cite{BeltranJimenez:2019tjy} and modifications thereof. Besides discussing its underlying mathematical concepts and giving a detailed account on its usage, we provided an example application to a simple scalar-tensor theory of gravity. We believe that this package will find wide application to assess the viability of gravity theories by comparing their post-Newtonian limit to observations in the solar system.

Despite the generality of \xPPN, being applicable to gravity theories based on different geometric foundations, various extensions and modifications are possible and may be implemented in future versions. For example, one may consider the following types of extensions:
\begin{enumerate}
\item
Alternative formulations of the PPN formalism: A newer approach towards the PPN formalism employs a different density variable~\cite[Sec. 4]{Will:2018bme}, which has the advantage that it simplifies the gauge transformation of the PPN potentials compared to the original formulation. Another approach to simplify the issue of gauge transformations is to resort to a formulation which makes use of gauge-invariant perturbation theory, thereby resolving the necessity of gauge considerations and simplifying the application of the PPN procedure~\cite{Hohmann:2019qgo}. Such modifications may be included by changing the expansion of the metric in terms of PPN potentials.

\item
Additional post-Newtonian potentials and parameters: Various theories of gravity exhibit a post-Newtonian limit in which the metric perturbation cannot be expressed only in terms of the post-Newtonian potentials used in the standard formalisms, and so more general potentials and corresponding parameters have been introduced. The presence of massive fields leads to the appearance of Yukawa-type potentials~\cite{Zaglauer:1990yh,Helbig:1991pk}, while higher-order derivatives can be accommodated by further integrals in the post-Newtonian potentials~\cite{Gladchenko:1990nw,Gladchenko:1994wu}. Another class of terms arises from theories which include parity-violating contributions~\cite{Alexander:2007zg,Alexander:2007vt,Alexander:2009tp}, or by violation of diffeomorphism invariance~\cite{Lin:2012bs,Lin:2012ea,Lin:2013tua}. Further, one may introduce also fourth-order tensor potentials to expand the term \(\order{g}{4}_{ab}\), which is neglected in the standard PPN formalism, but may be used to describe higher-order corrections to light propagation~\cite{Richter:1982zz,Richter:1982zza}. Such additional potentials may easily be included by defining the corresponding tensor fields and the equations which relate them to the corresponding source terms, in full analogy to the standard PPN potentials.

\item
Higher than fourth order in the post-Newtonian expansion: While in the standard PPN formalism implemented in the present version of \xPPN{} the metric (and possible other fields present in the theory) are expanded only up to the fourth velocity order, one may also consider higher order terms, and address problems such as the post-Newtonian dynamics of systems involved in the generation of gravitational waves~\cite{Blanchet:2013haa,Will:2018bme}. Care must be taken since the Euler equations governing the dynamics of the source matter attain dissipative terms at higher orders, and also the gravitational field exhibits dissipative behavior due to the loss of energy by emitted gravitational waves. Further, such as extension requires the definition of higher-order post-Newtonian potentials, along with corresponding post-Newtonian parameters, whose relation to observations and associated observables must be defined, so that such an extension would be a major task.

\item
Generalized post-Newtonian expansion: The standard PPN formalism assumes the validity of a post-Newtonian expansion around a flat Minkowski background. This assumption may be generalized, by allowing for a time-dependent Friedmann-Lemaître-Robertson-Walker background~\cite{Sanghai:2016tbi}, or by including non-perturbative effects arising from the Vainshtein screening mechanism~\cite{Avilez-Lopez:2015dja}.

\item
More general geometric frameworks: Another assumption of the standard PPN formalism states that the motion of test masses is governed by a single metric, and so the observable effects on test masses are fully covered by a post-Newtonian expansion of this single metric. This assumption may be relaxed if there are several dynamical metrics present in the considered gravity theory, which lead to more complex test matter dynamics and generalized sources of gravity~\cite{Clifton:2010hz,Hohmann:2010ni,Hohmann:2013oca}. Also more general connections may be considered, such as a general teleparallel connection~\cite{Jimenez:2019ghw}, as well as Riemann-Cartan~\cite{Smalley:1980em,Castagnino:1985,Castagnino:1987,Gladchenko:1990nw,Gladchenko:1994wu} or metric-affine geometry~\cite{Hehl:1994ue}. In theories of this class additional matter currents may appear from a coupling of spin to gravity, which gives rise to additional source terms, which must be accommodated for by further PPN potentials, together with their respective equations of motion and conservation laws generalizing the Euler equations~\cite{Weyssenhoff:1947iua,Ray:1982qr,Kopczynski:1986ep,Obukhov:1987yu,Obukhov:1993pt,Obukhov:1996mg,Puetzfeld:2014qba,Iosifidis:2020gth}.
\end{enumerate}

\begin{acknowledgments}
The author thanks Yuri Obukhov and Dirk Puetzfeld for helpful comments on the manuscript and pointing out related work. He gratefully acknowledges the full support by the Estonian Research Council through the Personal Research Funding project PRG356, as well as the European Regional Development Fund through the Center of Excellence TK133 ``The Dark Side of the Universe''.
\end{acknowledgments}

\appendix

\section{Implementation notes}\label{app:implementation}
Although \xTensor{} offers a possibility to split tensor indices for product manifolds and to project tensors to submanifolds, while \xPert{} provides an implementation of a perturbative expansion of tensor fields, these approaches are not the most well-adapted to the mathematical concepts discussed in section~\ref{sec:concepts}. While it may still be possible to use these packages for the task at hand, \xPPN{} provides its own implementations of these concepts, in order to match as closely as possible with the PPN formalism, and to simplify the different perturbative treatment of space and time components of tensors and in particular their derivatives. This appendix contains a few notes how the functions detailed in the main part of this article are implemented. The internal representation of split tensor fields is shown in section~\ref{app:intrep}. In section~\ref{app:31algo}, we give an overview of the algorithm used to obtain this decomposition for arbitrary tensor fields. Finally, in section~\ref{app:velalgo} we give a few notes on the implementation of the perturbative expansion in velocity orders. The aim of this technical appendix is to give an insight to the internal data structures and algorithms used by \xPPN, which is not necessary for its use, but may be helpful for readers who intend to develop extensions to \xPPN{} or similar packages based on \xAct{} in other areas of physics.

\subsection{Internal representation of decomposed tensors}\label{app:intrep}
As detailed in section~\ref{ssec:31split}, we adhere to the following mathematical interpretation of the $3+1$ split of tensor fields, which turns out to be convenient for the PPN formalism:
\begin{enumerate}
\item
The spacetime manifold is regarded as a product manifold \(M_4 \cong T_1 \times S_3\).
\item
The components of any tensor defined on the spacetime manifold \(M_4\) with respect to adapted coordinates \((t, x^a)\) are split into time and space components. For example, given a vector \(A^{\alpha}\), one has a decomposition into a time component \(A^0\) and spatial components \(A^a\).
\item
The decomposed parts of a tensor on \(M_4\) are regarded as tensors on \(S_3\), obtained as a time slice in a foliation of spacetime, carrying an additional dependence on time \(t\), where time is treated not as a coordinate, but as a parameter.
\item
The decomposition simplifies in the presence of tensor symmetries. For example, given an antisymmetric tensor \(A_{\alpha\beta} = A_{[\alpha\beta]}\), the component \(A_{00}\) vanishes, indices in \(A_{a0}\) may equivalently be sorted in canonical (lexicographic) order to yield \(-A_{0a}\) and \(A_{ab}\) is a tensor which inherits the antisymmetry in its two indices.
\end{enumerate}
The two mentioned approaches implemented in \xTensor{} do not match these criteria:
\begin{enumerate}
\item
The projection approach using normal vector fields and induced metrics does not perform any decomposition of indices; the equivalent of the time component \(A^0\) retains its tensor rank, and is hence represented by a vector, although being projected onto (and hence tangent to) a one-dimensional manifold. One may obtain a scalar by contracting with the normal vector field, at the cost of computational complexity for carrying this contracted factor through all calculations\footnote{This approach is used in \xPand{} \cite{Pitrou:2013hga}, and such contractions are then replaced by tensors of lower rank which are pre-defined for the space and time components of metric perturbations. Here we aim to avoid manually defining such a decomposition, in favor of an algorithm which performs the decomposition automatically whenever a tensor field is defined.}.
\item
The split along product manifolds also retains the tensor rank. Instead of the inert time index \(0\) and thus being treated as scalar, time components carry a coordinate index associated to the time manifold, which therefore must satisfy the rules imposed by \xTensor{} on such indices (such as the restriction on the number of occurrences, uniqueness in expressions, summation over dummy indices etc.).
\end{enumerate}
Furthermore, in both approaches time retains its nature of a coordinate. Therefore, \xPPN{} uses the following approach to decomposing tensor fields instead. The components of the $3+1$ split of tensors are represented by the inert function \lstinline/PPNTensor/ in \xPPN, which resides in the private context \lstinline/xAct`xPPN`Private`/. It may be used in the following two forms:
\begin{enumerate}
\item
\lstinline/PPNTensor[$h$, {$s_1$, $s_2$, $\ldots$, $s_k$}]/, where \(h\) is a valid tensor head belonging to a tensor on the spacetime manifold \(M_4\) and \(s_1, s_2, \ldots, s_k\) is a list of index slots, represents a component in the $3+1$ decomposition of the tensor with head \(h\). The particular component is selected by the list of tensor slots; see the list below for valid values and their interpretation.
\item
\lstinline/PPNTensor[$h$, {$s_1$, $s_2$, $\ldots$, $s_k$}, $n$]/, where \(n\) is a non-negative integer and \(h, s_1, s_2, \ldots, s_k\) are as above, represents the term of velocity order \(\mathcal{O}(n)\) in the perturbative expansion of the aforementioned component represented by \lstinline/PPNTensor[$h$, {$s_1$, $s_2$, $\ldots$, $s_k$}]/.
\end{enumerate}
The valid values for the slots \(s_1, s_2, \ldots, s_k\) depends on the slots \(S_1, S_2, \ldots, S_k\) of \(h\). The number $k$ of slots given (i.e., the length of the list) must match the number of slots of \(h\). Further, for every slot \(S_i\) of \(h\) the following rules must be satisfied:
\begin{enumerate}
\item
The character of the slots must match, i.e., if \(S_i\) is an upper (lower) index, then also \(s_i\) must be an upper (lower) index.
\item
If \(S_i\) is \lstinline/$\pm$TangentMfSpacetime/, then \(s_i\) must be one of \lstinline/$\pm$Labels/ or \lstinline/$\pm$TangentMfSpace/.
\item
If \(S_i\) is \lstinline/$\pm$LorentzMfSpacetime/, then \(s_i\) must be one of \lstinline/$\pm$Labels/ or \lstinline/$\pm$LorentzMfSpace/.
\item
If \(S_i\) is any other slot and hence does not belong to a vector bundle over \(M_4\), then \(s_i\) must be the same as \(S_i\).
\end{enumerate}
The interpretation should be evident: every tangent or Lorentz index associated to the spacetime manifold \(M_4\) is replaced either by a corresponding index on the space manifold or the special inert value \lstinline/LI[0]/, belonging to \lstinline/Labels/, which represents the time component. \xPPN{} defines upvalues for the following \xTensor{} functions applied to \lstinline/PPNTensor/ objects obeying any of the two forms listed above:
\begin{enumerate}
\item
\lstinline/xTensorQ/ yields \lstinline/True/ if $h$ is a valid tensor head on the spacetime manifold and the slots satisfy the conditions listed above, and \lstinline/False/ otherwise.
\item
\lstinline/SlotsOfTensor/ returns the list \lstinline/{$s_1$, $s_2$, $\ldots$, $s_k$}/ of tensor slots.
\item
\lstinline/DependenciesOfTensor/ gives the dependencies of \(h\), but with \lstinline/MfSpacetime/ replaced by \lstinline/MfSpace/ and \lstinline/TimePar/. If the tensor depends on any other parameters or manifolds, then these dependencies are preserved.
\item
\lstinline/SymmetryGroupOfTensor/ returns the remaining symmetry group of the $3+1$ decomposed tensor.
\item
\lstinline/PrintAs/ yields the same symbol \lstinline/PrintAs[$h$]/ which is used for printing \(h\) if no velocity order is given, and otherwise adds the velocity order \(n\) as an overscript \(\order{h}{n}\).
\end{enumerate}
Further, \lstinline/PPNTensor/ automatically applies rules which are obtained from index symmetries. These are calculated whenever a tensor field is defined, as explained below.

\subsection{$3+1$ spacetime split algorithm}\label{app:31algo}
In order to retain only independent components in the $3+1$ split of a tensor, \xPPN examines tensor symmetries whenever a new tensor on the spacetime manifold is defined. It does so by extending the \xTensor{} function \lstinline/DefTensor/, using the extensibility framework implemented in \xAct, to perform the following steps:
\begin{enumerate}
\item
Determine which index slots of the tensor are associated to the spacetime manifold. If a tensor carries additional, internal indices, then they will be treated separately and ignored in the following steps.
\item
Calculate the symmetry group acting on the spacetime indices only, i.e., factor out any possible symmetries acting on internal indices and also ignore them in the following steps.
\item
Create a list of all possible ways to split the spacetime indices into space and time components.
\item
Consider the action of the symmetry group on the splits from the previous step and determine the orbits of this action. For example, if a tensor carries a symmetry (or antisymmetry) in two indices, then there is no difference between choosing the first index to be temporal and the second index spatial or vice versa, and so these two possible splits belong to the same orbit; these two possible ways to arrange the indices of the split tensor are equivalent.
\item
For each orbit, perform the following steps:
\begin{enumerate}
\item
Pick a representative, i.e., from all equivalent ways to arrange the indices, choose the one which comes first in canonical (lexicographic) order.
\item
Check whether any of the elements of the subgroup which permutes only the temporal indices comes with an antisymmetry, i.e., changes the sign of the tensor. If this is the case, this component must vanish identically, since permuting only temporal indices must leave the tensor invariant, and so it must be equal to its negative. In this case, define all index combinations in this orbit to be an alias for the zero tensor \lstinline/Zero/.
\item
If the chosen representative does not vanish, conclude with the following steps:
\begin{enumerate}
\item
Determine the remaining symmetry group acting on the spatial indices only.
\item
Define a tensor on a purely spatial manifold, whose temporal and spatial indices match with the representative, which carries the symmetry determined in the previous step in the spacetime indices and any inherit symmetry in internal indices, and which in addition depends on a time parameter.
\item
Define all other arrangements of temporal and spatial indices which are equivalent to the representative as alias for the previously defined tensor or its negative, taking into account possible sign changes when indices are permuted.
\end{enumerate}
\end{enumerate}
\end{enumerate}
To illustrate these steps, consider the case of defining an antisymmetric tensor \(A_{\alpha\beta} = A_{[\alpha\beta]}\) on the spacetime manifold, by invoking
\begin{lstlisting}
In[]:= DefTensor[A[-T4<a>, -T4<b>], MfSpacetime, Antisymmetric[{1, 2}]]
\end{lstlisting}
Then the following steps are carried out automatically by \xPPN:
\begin{enumerate}
\item
There are no internal indices, and so the spacetime related symmetry operations acting on the tensor indices are the identity and the swapping of the two indices, where the latter also changes the sign of the tensor.
\item
Each of the two indices of \(A_{\alpha\beta}\) splits into time and space. Hence, there are four possible decomposed tensor fields: \(A_{00}, A_{0a}, A_{a0}, A_{ab}\).
\item
The two components \(A_{0a}\) and \(A_{a0}\) are transformed into each other under the index symmetry of the original tensor fields; hence, they belong to the same orbit of the action of the symmetry group. The remaining components are the sole elements in their respective orbits.
\item
The three orbits are examined separately:
\begin{enumerate}
\item
For \(A_{00}\) one finds that the swap operation acts on temporal indices only, but also changes the sign of the tensor. Hence, \(A_{00} = -A_{00} = 0\) and the tensor is defined as an alias of \lstinline/Zero/:
\begin{lstlisting}
A /: PPNTensor[A, {-Labels, -Labels}] := Zero;
A /: PPNTensor[A, {-Labels, -Labels}, _] := Zero;
\end{lstlisting}
\item
For \(A_{0a}\) and \(A_{a0}\) the former arrangement of indices is chosen as a representative, since the indices are in canonical (lexicographic) order. A purely spatial tensor \(A_{0a}\) is defined, which has one free index slot. \(A_{a0}\) is defined as an alias for \(-A_{0a}\):
\begin{lstlisting}
A /: PPNTensor[A, {-TangentMfSpace, -Labels}] :=
	(-PPNTensor[A, {-Labels, -TangentMfSpace}][#2, #1] &);
\end{lstlisting}
\item
Another spatial tensor \(A_{ab}\) is defined, which is antisymmetric in its two index slots. The tensor symmetry is again associated to the symbol \lstinline/A/.
\end{enumerate}
\end{enumerate}
Note that all properties of these tensors, such as their symmetries and tensor slots, are associated with the symbol (tensor head) which is used in the original call to \lstinline/DefTensor/ to define the tensor on the spacetime manifold, and no further symbols are introduced.

\subsection{Velocity order decomposition algorithm}\label{app:velalgo}
In order to implement the rules for the perturbative expansion in velocity orders detailed in section~\ref{ssec:perturb}, a few special cases have been defined for the function \lstinline/VelocityOrder/ shown in section~\ref{ssec:veldecomp}. For the product rule~\eqref{eq:velordprod}, it is most convenient to rely on Mathematica's pattern matching and recursive function application functionality. Given a product \(A = A_1 \cdots A_N\) of \(N\) tensors, one may split off the first factor,
\begin{equation}
A = A_1\tilde{A}\,, \quad \tilde{A} = A_2 \cdots A_N\,,
\end{equation}
and then recursively apply the rule
\begin{equation}
\order{A}{n} = \sum_{k = 0}^{n}\order{A}{k}_1\order{\tilde{A}}{n - k}\,.
\end{equation}
This is repeated until only a single factor is left. The implementation then takes the following simple form.
\begin{lstlisting}
VelocityOrder[Times[ex0_, ex1__], n_, opt : OptionsPattern[]] :=
	Module[{k}, Sum[VelocityOrder[ex0, k, opt] *
		VelocityOrder[Times[ex1], n - k, opt], {k, 0, n}]];
\end{lstlisting}
Next, we come to the relation~\eqref{eq:velordfunc} for the perturbative expansion of expressions which are given as functions of an arbitrary number of arguments. To evaluate this formula, it is useful to define the formal series
\begin{equation}
\tilde{A}_i(\epsilon) = \sum_{k = 0}^{\infty}\epsilon^k\order{A}{k}_i
\end{equation}
for the tensor fields \(A_1, \ldots, A_N\). Observe that the $n$'th order series coefficient
\begin{equation}
\frac{1}{n!}\left.\frac{\dd^n}{\dd\epsilon^n}f(\tilde{A}_1(\epsilon), \ldots, \tilde{A}_N(\epsilon))\right|_{\epsilon = 0} = \sum_{p_{ik}}f^{(p_{11} + \ldots + p_{1n}, \ldots, p_{N1} + \ldots + p_{Nn})}\left(\order{A}{0}_1, \ldots, \order{A}{0}_N\right)\prod_{i = 1}^{N}\prod_{k = 1}^{n}\frac{\order{A}{k}_i^{p_{ik}}}{p_{ik}!}\,,
\end{equation}
where the sum runs over
\begin{equation}
p_{ik} \in \{0, \ldots, q_k\}\,, \quad
\sum_{i = 1}^{N}p_{ik} = q_k\,; \quad
i \in \{1, \ldots, N\}
\end{equation}
and
\begin{equation}
q_k \in \{1, \ldots, n\}\,, \quad
\sum_{k = 1}^{n}kq_k = n\,,
\end{equation}
is exactly the term of $n$'th velocity order in the expansion~\eqref{eq:velordfunc}. Hence, it can be obtained as follows.
\begin{lstlisting}
VelocityOrder[fkt_ ? ScalarFunctionQ[args__], n_, opt : OptionsPattern[]] :=
	Module[{t, i}, SeriesCoefficient[
		fkt @@ (Sum[t^i * VelocityOrder[#, i, opt], {i, 0, n}] & /@ {args}),
	{t, 0, n}]];
\end{lstlisting}
Finally, we have encountered the rule~\eqref{eq:velordtimed} that time derivatives are weighted with an additional velocity order. Since time derivatives are represented by parameter derivatives with respect to \lstinline/TimePar/, this behavior is achieved by counting time derivatives as follows.
\begin{lstlisting}
VelocityOrder[ParamD[t : TimePar ..][expr_], n_, opt : OptionsPattern[]] :=
	ParamD[t][VelocityOrder[expr, n - Length[{t}], opt]];
\end{lstlisting}
Note that partial derivatives with respect to spatial coordinates do not incur additional velocity orders.
\begin{lstlisting}
VelocityOrder[PD[i_][expr_], n_, opt : OptionsPattern[]] :=
	PD[i][VelocityOrder[expr, n, opt]];
\end{lstlisting}

\bibliography{xppn}
\end{document}